\def\inarxiv{}
\documentclass[10pt,letterpaper,twocolumn,table]{article}
\usepackage{usenix-2020-09}

\usepackage[square,comma,numbers,sort&compress]{natbib}

\usepackage{amssymb}
\usepackage{amsmath,amsopn, amsfonts}
\usepackage{caption}
\usepackage{subcaption}
\usepackage{endnotes,microtype,xspace,graphicx,fancyvrb,multirow}
\usepackage{booktabs}
\usepackage{array,underscore,relsize}
\usepackage[T1]{fontenc}
\usepackage{times}
\usepackage{fancyhdr}

\usepackage[inline]{enumitem}
\usepackage[labelfont=bf,font=small,skip=5pt]{caption}

\usepackage{fp}
\usepackage{siunitx}

\usepackage{algorithm}
\usepackage[noend]{algpseudocode}
\algrenewcommand\algorithmicindent{0.75em}
\usepackage{balance}

\usepackage[super]{nth}
\usepackage{pifont}
\usepackage{import}
\usepackage[normalem]{ulem}

\usepackage{placeins}
\usepackage{longtable}
\usepackage{supertabular}
\usepackage{xr}

\usepackage{pgffor}

\newcommand{\sys}{\mbox{\cc{autofz}}\xspace}
\newcommand{\sysn}{\mbox{\cn{autofz}}\xspace}
\newcommand{\sysm}{\mbox{\cc{autofz}\textsuperscript{-}}\xspace}

\newcommand{\cc}[1]{\mbox{\smaller[0.5]\texttt{#1}}}
\newcommand{\cn}[1]{\mbox{\texttt{#1}}}

\fvset{fontsize=\scriptsize,xleftmargin=8pt,numbers=left,numbersep=5pt}

\input{code/fmt}

\def\Snospace~{\S{}}

\newif\ifdraft\drafttrue
\newif\ifnotes\notestrue
\ifdraft\else\notesfalse\fi

\input{glyphtounicode}
\newcolumntype{R}[1]{>{\raggedleft\let\newline\\\arraybackslash\hspace{0pt}}p{#1}}

\newcommand{\squishlist}{
\begin{itemize}[noitemsep,nolistsep]
  \setlength{\itemsep}{-0pt}
}
\newcommand{\squishend}{
  \end{itemize}
}

\usepackage{tikz}
\newcommand*\WC[1]{%
\begin{tikzpicture}[baseline=(C.base)]
\node[draw,circle,inner sep=0.2pt](C) {#1};
\end{tikzpicture}}

\usepackage{xstring}
\newcommand{\PP}[1]{
\vspace{2px}
\noindent{\bf \IfEndWith{#1}{.}{#1}{#1.}}
}

\newcommand{\PN}[1]{
\vspace{2px}
\noindent{\bf #1}
}

\newcommand{\PPI}[1]{
\noindent{\bf \IfEndWith{#1}{.}{#1}{#1.}}
}

\newcommand{\PNI}[1]{
\noindent{\bf #1}
}

\newcommand{\GB}{\,\text{GB}\xspace}

\newcommand{\boxbeg}{
\vspace{2px}
\noindent\begin{tabular}{|l|}\hline
\begin{minipage}{3.2in}
\vspace{2px}
\noindent
}

\newcommand{\boxend}{
\vspace{2px}
\end{minipage}\\ \hline
\end{tabular}
\vspace{-10pt}
}

\newcommand{\enfuzz}{\textsc{EnFuzz}\xspace}

\newcommand{\cupid}{\textsc{Cupid}\xspace}

\newcommand{\collabfuzz}{\textsc{CollabFuzz}\xspace}

\newcommand{\afl}{AFL\xspace}
\newcommand{\AFL}{AFL\xspace}
\newcommand{\aflfast}{AFLFast\xspace}
\newcommand{\mopt}{MOpt\xspace}
\newcommand{\qsym}{QSYM\xspace}
\newcommand{\angora}{Angora\xspace}

\newcommand{\lafintel}{\textsc{LAF-Intel}\xspace}
\newcommand{\radamsa}{Radamsa\xspace}
\newcommand{\redqueen}{Redqueen\xspace}
\newcommand{\libfuzzer}{libFuzzer\xspace}
\newcommand{\fairfuzz}{FairFuzz\xspace}
\newcommand{\learnafl}{LearnAFL\xspace}

\newcommand{\unifuzz}{\textsc{UniFuzz}\xspace}
\newcommand{\ftsFull}{Fuzzer Test Suite\xspace}
\newcommand{\fts}{FTS\xspace}

\newcommand{\libarchive}{\cc{libarchive}\xspace}
\newcommand{\pdftotext}{\cc{pdftotext}\xspace}
\newcommand{\boringssl}{\cc{boringssl}\xspace}

\newcommand{\freetype}[1]{\cc{freetype2}\xspace}
\newcommand{\woff}[1]{\cc{woff2}\xspace}
\newcommand{\ffmpeg}{\cc{ffmpeg}\xspace}

\newcommand{\exiv}[1]{\cc{exiv2}\xspace}

\newcommand{\re}[1]{\cc{re2}\xspace}

\newcommand{\nm}{\cc{nm}\xspace}
\newcommand{\mujs}{\cc{mujs}\xspace}
\newcommand{\tcpdump}{\cc{tcpdump}\xspace}

\newcommand{\prep}{T_{prep}}
\newcommand{\focus}{T_{focus}}

\newcounter{rowno}
\definecolor{ForestGreen}{RGB}{34,139,34}
\newcommand{\cmark}{\color{ForestGreen}\ding{51}}
\newcommand{\xmark}{\color{red}\ding{55}}
\newcommand{\trimark}{\color{brown}\ding{115}}

\newcommand{\pdiff}{diff_{peak}}

\newcommand{\eautoref}[1]{\ifdefined\inarxiv\autoref{#1}\else\autoref*{#1} in \cite{autofz-extended}\fi}

\newcommand{\onlymain}[1] {\ifdefined\inarxiv{}\else{#1}\fi}

\makeatletter

\newcommand*{\addFileDependency}[1]{%
\typeout{(#1)}%
\@addtofilelist{#1}
\IfFileExists{#1}{}{\typeout{No file #1.}}
}\makeatother

\newcommand{\ignore}[2]{\hspace{0in}#2}

\newcommand{\replacex}[1]{%
 \StrLen{#1}[\stringlength]%
  \newcount\loopcounter
  \loopcounter=0
  \loop\ifnum\loopcounter<\stringlength%
    x%
    \advance\loopcounter by 1%
  \repeat%
}

\gdef\therev{}
\gdef\thedate{}

\begin{document}

\title{\sys: Automated Fuzzer Composition at Runtime}

\ifdefined\DRAFT
 \pagestyle{fancyplain}
 \lhead{Rev.~\therev}
 \rhead{\thedate}
 \cfoot{\thepage\ of \pageref{LastPage}}
\fi

\author{
 {\rm Yu-Fu Fu \hspace{1em} Jaehyuk Lee \hspace{1em} Taesoo Kim} \\\\
 Georgia Institute of Technology %
}

\maketitle

\begin{abstract}
Fuzzing has gained in popularity for
software
vulnerability detection
by virtue of the tremendous effort
to develop a diverse set of fuzzers.
Thanks to various fuzzing techniques,
most of the fuzzers
have been able to demonstrate
great performance on their selected targets.
However,
paradoxically,
this diversity in fuzzers
also made it difficult
to select fuzzers that are
best suitable for complex real-world programs,
which we call \emph{selection burden}.
Communities attempted to address this problem
by creating a set of standard benchmarks
to compare and contrast the performance
of fuzzers
for a wide range of applications,
but the result was always a suboptimal decision---%
the best performing fuzzer on \emph{average}
does not guarantee the best outcome
for the target of a user's interest.

To overcome this problem,
we propose an automated, yet non-intrusive meta-fuzzer,
called \sys\footnote{\sys is available at \href{https://github.com/sslab-gatech/autofz}{https://github.com/sslab-gatech/autofz}},
to maximize the benefits of existing state-of-the-art fuzzers via \emph{dynamic composition}.
To an end user,
this means that, instead of spending time on selecting which fuzzer to adopt
(similar in concept to hyperparameter tuning in ML),
one can simply put \emph{all} of the available fuzzers to \sys
(similar in concept to AutoML),
and achieve the best, optimal result.
The key idea is
to monitor the runtime progress of the fuzzers, called trends
(similar in concept to gradient descent),
and make a fine-grained adjustment of resource allocation (e.g., CPU time) of each fuzzer.
This is a stark contrast to existing approaches that statically combine a set of fuzzers,
or via exhaustive pre-training \emph{per target program}---%
\sys deduces a suitable set of fuzzers of the \emph{active workload}
in a fine-grained manner at runtime.
Our evaluation shows that,
given the same amount of computation resources,
\sys
outperforms
any best-performing individual fuzzers in 11 out of 12 available benchmarks
and beats the best, collaborative fuzzing approaches
in 19 out of 20 benchmarks
without any prior knowledge in terms of coverage.
Moreover, on average, \sys found 152\% more bugs than individual fuzzers on \unifuzz and \fts, and 415\% more bugs than collaborative fuzzing on \unifuzz.

\end{abstract}

\date{}

\sloppy

\section{Introduction}

Complications in modern programs
inevitably entangle the manual analysis of software
and further reduce the chance of discovering software defects.
To overcome this, researchers and industry
have conducted extensive studies on
discovering software vulnerabilities with automation, called fuzzing~\cite{miller1990, boehme21:fuzzing-challenge, manes:tse:2021}.
Fuzzing generates input expected to trigger the defects in the program
with feedback retrieved from multiple rounds of execution
and monitors for any anomalies.
To enhance the performance of fuzzing,
a great deal of research using different techniques
has been proposed%
~\cite{driller-stephens-ndss16, yun18:qsym,
chen2018angora, cadar:klee08, vuzzer-rawat-ndss17,
langfuzz-holler-sec12,cab-fuzz-kim-atc17, cgf-bohme-ccs16,
xu:os-fuzz, swiecki21:honggfuzz, afl-zalewski-web, wen20:memlock, petsios17:slowfuzz, lemieux18:fairfuzz, learnafl,
fioraldi20:afl++, aschermann19:redqueen, lyu19:mopt, radamsa, lafintel, bohme20:entropy}.
As a result, fuzzing has demonstrated its effectiveness
in disclosing vulnerabilities
from off-the-shelf binaries%
~\cite{ding21:empir-study-oss-fuzz-bugs, afl-found-cve,
afl-zalewski-web, syzkaller-found-cve, honggfuzz-found-cve}.
Motivated by convincing empirical evidence
for the practicality of fuzzing,
Google~\cite{oss-fuzz-google-web,clusterfuzz-google-web} and,
recently, Microsoft~\cite{microsoft-onefuzz}
have deployed scalable fuzzers.

\label{s:intro}
\begin{figure}[t]
  \begin{center}
   {
    \fontsize{6pt}{6pt}\selectfont
     \def\svgwidth{\columnwidth}
     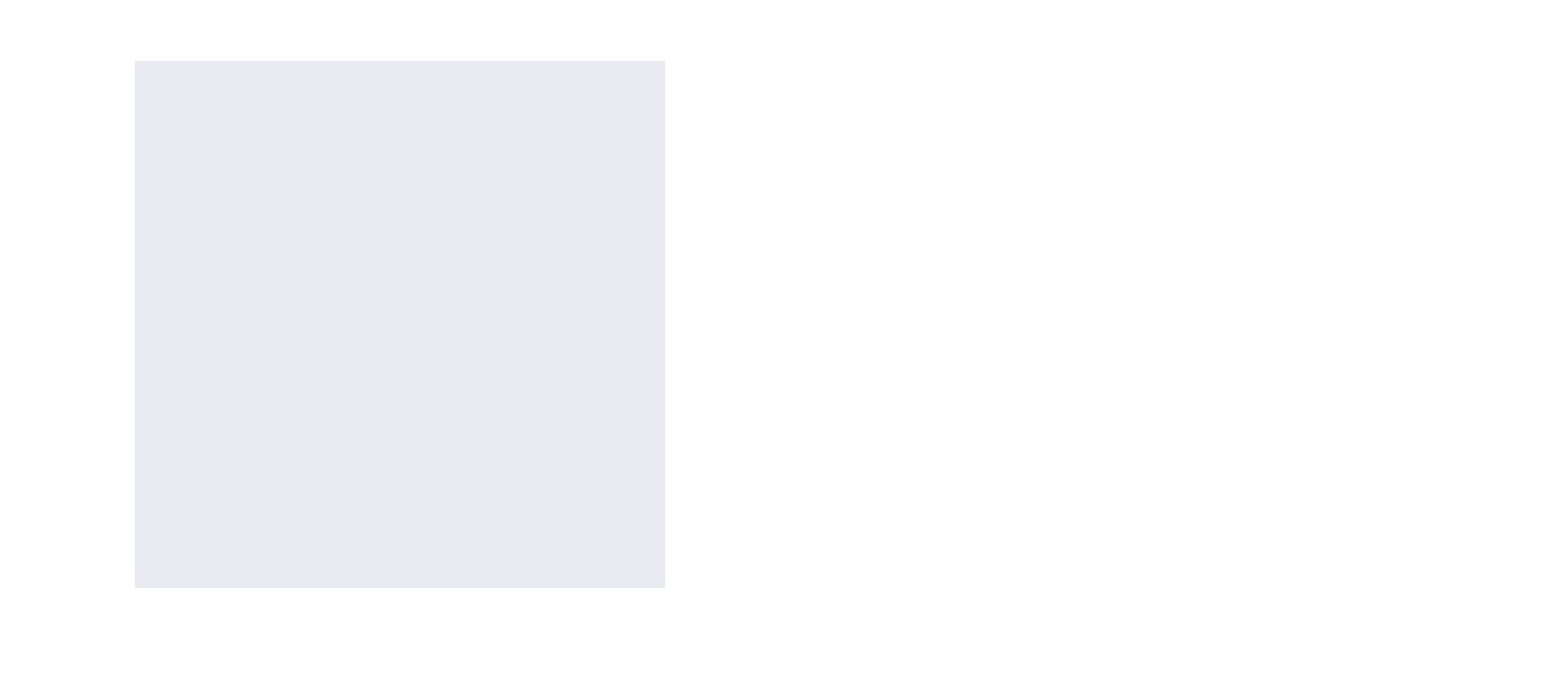
    }
  \end{center}
  \caption{Performance of \sys and baseline fuzzers
    presented with coverage ratio of fuzzing \cc{exiv2} and \cc{ffmpeg} during 24 hours.
    Each graph is generated with an arithmetic mean and 80\% confidence interval for 10 fuzzing rounds.
    Coverage ratio is a percentage of branches explored by each fuzzer.
    We carefully selected two test suites from \autoref{f:autofuzz-eval-best}
    to highlight the motivation of \sys.
    }
  \label{f:motivation}
\end{figure}

However, the remarkable enhancement and diversity of fuzzers
create a \emph{selection burden}, paradoxically,
that requires another significant engineering effort
to select the best-performing fuzzer(s) per target
(similar in concept to hyperparameter tuning).
Note that fuzzer selection is a dominant factor
that contributes to vulnerability detection and its efficiency.
As shown in~\autoref{f:motivation},
the outcome significantly varies depending on
which fuzzer is selected.
For example, with the same amount of resources,
\redqueen outperforms \radamsa
by more than 10 times in fuzzing \exiv2
(left graph).

A handful of research efforts have tried to mitigate the selection problem \cite{klees18:evaluat-fuzz-testin}.
Fuzzing benchmarks~\cite{darpa:cgc, fts, fuzzbench,
  dolan-gavitt16:lava,hazimeh21:magma, li21:unifuzz, profuzzbench}
enumerate well-suited fuzzers for each benchmark target.
The evaluation of the benchmark
helps users to understand which fuzzers are favorable
for fuzzing different types of binaries.
Recently, collaborative fuzzing%
~\cite{chen19:enfuz, guler20:cupid, osterlund21:collab}
has showcased that
cooperating different combinations of fuzzers
sometimes outperforms individual fuzzers
thanks to their corpus sharing.
However, this still imposes the burden of selecting
what fuzzers to put into an ensemble.
Moreover, the selection of fuzzers
generally requires
significant computing and human resources
because it relies on static information.

Unlike previous works,
\sys automatically deploys a set of fuzzer(s)
\emph{per workload, not per program}.
The goal of \sys is
to completely automate the selection problem
via the \emph{dynamic composition} of fuzzers
as a push-button solution.
Therefore,
when end users select a set of baseline fuzzers,
\sys automatically devises the best performance
utilizing runtime information.

\label{ss:intro:challenge}
\noindent
\sys is largely motivated by the following observations:

\PP{1) No universal fuzzer invariably outperforms others.}
As shown in ~\autoref{f:motivation},
a particular fuzzer cannot persistently achieve optimal performance
independent of the workload.
For example, \learnafl is demonstrated
to be the best-performing fuzzer
for fuzzing the \ffmpeg binary.
However, when the target is changed to \exiv2,
\learnafl is relegated to sixth place,
which shows the lack of consistency in the performance of fuzzers
against different binaries.
Unfortunately,
this inconsistency is not uncommon~\cite{li21:unifuzz,fuzzbench}.
Therefore,
to achieve better performance,
significant engineering effort is required
to handpick the fuzzers
whenever the target binary is changed
or a new fuzzer is introduced.

\PP{2) The efficiency of each fuzzer is not perpetual throughout.}
As shown in~\autoref{f:motivation},
initially, \angora significantly outperforms the others
in fuzzing the \exiv2 binary.
However, \lafintel and \redqueen
come from behind and take the lead after about two hours.
We call this rank inversion.
Note that
another inversion occurs after 12 hours have elapsed.
\lafintel initially slightly outperforms \redqueen,
but \redqueen becomes the best
in the end result.
Moreover, we found that rank inversion occurs more frequently
when seed synchronization is introduced
to share the interesting input among multiple fuzzers (\autoref{ss:eval-decision}).
This result indicates that
sticking to the initially chosen fuzzer
during the entire execution
cannot achieve optimal performance.

\PP{3) Naive resource allocation results in inefficiency.}
Unfortunately,
all collaborative fuzzing approaches
have focused only on
selecting the best combination of fuzzers and
running the fuzzers
assigned with equally partitioned resources.
However,
in addition to fuzzer selection,
the resource distribution among the selected fuzzers
can severely affect the performance of fuzzing.
To achieve better results,
the performance of each fuzzer should be assessed
and considered to distribute resources
among the selected fuzzers.

\PP{4) Randomness in fuzzing prevents reproductions}
More important, note that this trend
will not invariably be captured
in different fuzzing executions
because fuzzing is inherently a random process.
Therefore, even though experts
put a lot of effort into
locating the best combinations of fuzzers and
associated guidance information such as
appropriate time to switch fuzzers and resource allocation,
this statically analyzed information
may not be consistent
across subsequent fuzzing runs,
even for the same target binary.
Consequently,
statically prearranging those configurations
might result in losing the benefits of
promptly dealing with a different workload
with most suitable fuzzers.

\section{Our Approach: Using Trend Per Workload}
Given the aforementioned challenges,
timely locating the best performing fuzzer(s) is non-trivial,
as the workload changes during the fuzzing execution.
In this paper,
we propose \sys, an automated
\emph{meta}-fuzzer
that outperforms the best individual fuzzers in any target
\emph{without} implementing any particular fuzzing algorithm.
The key idea of \sys
is to dynamically deploy a set of, or possibly all, of fuzzers
along with efficient utilization guidance,
such as resource allocation,
per workload, not per program.
We call the runtime progress of baseline fuzzers a \emph{trend}.
Specifically,
\sys switches fuzzers and adjusts resources
as the trend changes,
instead of adhering to a particular set of fuzzers,
during an entire execution.

\sys splits its execution into two different phases,
preparation and focus phases (\autoref{f:autofuzz-workflow}),
to monitor the trend changes.
The preparation phase captures the runtime trend of
target binaries and deploys fuzzers
that illustrate strong trends.
Based on the captured trend and the guidance information,
the focus phase tries to achieve the optimal performance
with the selected fuzzers.
Also, the dynamic resource adjustment of \sys
utilizing guidance information
allows it
to enjoy the best of both an individual fuzzer
and a combination of different fuzzers.
On the one hand,
it can prioritize a particular fuzzer,
significantly outperforming others
by allocating all resources to the selected fuzzer.
On the other hand,
\sys takes advantage of multiple fuzzers
by distributing resources.

More important,
unlike previous approaches
that utilize benchmark and offline analysis
to select well-suited fuzzers beforehand,
\sys does not require significant engineering effort
or hindsight
because it dynamically adopts the trends
and
automatically configures the best fuzzer set at runtime.
Therefore,
\sys can devote additional resources
that were previously spent on offline analysis
to the actual fuzzing campaign.
Moreover, such an approach bridges the gap
between developing a fuzzer and its utilization
and provides unprecedented opportunities. %
For example, even non-expert users,
without any knowledge about the selected fuzzers and target binaries,
can achieve better fuzzing performance
on any target program.
Furthermore,
we integrate \sys into \ftsFull and \unifuzz,
which support the most efficient and widely used fuzzers.
Therefore,
\sys can readily benefit from any fuzzer
that will be integrated into those benchmarks.

We evaluate \sys on \ftsFull and \unifuzz
to demonstrate how \sys can efficiently
utilize multiple different fuzzers.
Our evaluation demonstrates that \sys can significantly outperform
most of the individual fuzzers supported by the benchmarks
regardless of the target binaries.
Moreover, we expand \sys to support a multi-core environment
and compare it with collaborative fuzzing,
such as \enfuzz and \cupid.
We found that the resource allocation strategy and
fine-grained fuzzer scheduling of \sys
help it to achieve better performance
in most of the target binaries.

\noindent This paper makes the following main contributions:

\begin{itemize}
	\item \PP{A dynamic fuzzer composition per workload.}
	Existing approaches
	aim to find a static group of fuzzers
	that work best for each \emph{target program}.
	However,
	careful consideration of
	per-workload dynamic trends
	allows \sys to avoid
	benchmark-based decisions
	biased to a specific target program.
		Therefore,
		\sys does not need to utilize
		one static group of fuzzers
		during the entire fuzzing campaign.
		In essence,
		as well as consistently selecting well-performing fuzzers,
		\sys can give a second chance to fuzzers that have not been selected
		but turn out to be suitable for a particular workload later.

	\item \PP{Automatic and non-intrusive approach.}
	For end users,
	\sys is a push-button solution
	that automatically selects the best suitable fuzzers
	for any given target.
	For fuzzer developers,
	each fuzzer can be
	integrated to \sys
	with minimal engineering effort:
    126 LoC changes on average
    are required in the 11 fuzzers
    we support initially (\autoref{s:impl}).
	We call \sys a \emph{meta}-fuzzer
	because it implements no fuzzing algorithm internally.

	\item \PP{Efficient resource scheduling algorithm.}
		\sys measures the performance of individual fuzzers
		and efficiently distributes computational resources to
		the selected fuzzers.
		This helps \sys take advantage of
		the strong trend of individual fuzzers
		while maximizing the collaboration effects
		of multiple different fuzzers.
		Moreover,
		\sys first highlights the resource scheduling
		as another important factor affecting
		the efficiency of the collaboration.
		The collaboration effect can be maximized
		when the resources are properly assigned
		among the selected fuzzers.

\end{itemize}

\section{Related work}
\label{s:relwk}

\PP{Fuzzing benchmark}
The lack of metrics and representative target programs
in fuzzing research 
prevents their results from being reproducible,
making it difficult for users to decide which fuzzers 
are suitable for fuzzing specific target programs.
To mitigate this problem and provide a standardized 
test suite for various fuzzers, 
several fuzzing benchmarks
have been proposed.
LAVA-M~\cite{dolan-gavitt16:lava} provides a set of benchmark programs 
that contain syntactically injected out-of-bounds memory access vulnerabilities.
The Cyber Grand Challenge~\cite{darpa:cgc} benchmark
consists of various target binaries having a wide variety of synthetic software defects. 
\ftsFull (FTS) \cite{fts} evaluates 
fuzzers against real-world vulnerabilities.  
FuzzBench \cite{fuzzbench} improved FTS and
provide the frontend interface 
that allows smooth integration of new benchmarks and fuzzers. 
Moreover,
\unifuzz~\cite{li21:unifuzz} and Magma~\cite{hazimeh21:magma} provide
benchmarks that have real-world vulnerabilities   
together with their ground truth 
to verify genuine software defects   
from random software crashes.

\PP{Collaborative fuzzing.}
Collaborative fuzzing
attempts to improve the performance of a fuzzing campaign 
by orchestrating multiple different types of fuzzers
with seed synchronization (see below). 
\enfuzz~\cite{chen19:enfuz} first demonstrated that
deploying various types of fuzzers together
allows it to achieve better code coverage.
Recently, \cupid~\cite{guler20:cupid} showcased that 
offline analysis with a training set, 
including empirically collected representative branches,
is able to predict target-independent fuzzer combinations.
In addition to fuzzer selection,
\collabfuzz~\cite{osterlund21:collab} illustrates 
the importance of test case scheduling policies in seed synchronization among the selected fuzzers.
\begin{table}[h]
  \begin{center}
  \resizebox{\columnwidth}{!}{
    \begin{tabular}{lrrr}
    \toprule
                                    &  \enfuzz  & \cupid & \sys \\
    \midrule

      Number of selected fuzzers       &  user-configured  & user-configured & automatic \\
                                    &   (2-4) & (2-4) & (1-11) \\
      Fuzzer switches at runtime?   &    \xmark & \xmark & \cmark \\
      
      Require prior knowledge?      & \cmark  & \cmark & \xmark \\
      Cost of pre-training          & low  & high & none \\
      Target-independent decision?  & \xmark & \trimark & \cmark \\
      Cost of adding new fuzzers?   & \cmark  & \cmark & \xmark \\

      Resource allocation           & static & static & dynamic  \\
      Resource distribution policy  & equal & equal & proportional  \\
      \bottomrule
    
    \end{tabular}
    }
  \end{center}
  \caption{
    Comparison of \enfuzz, \cupid, and \sys.
  }
  \label{t:relwk-compare-sys}
\end{table}

\autoref{t:relwk-compare-sys} provides 
comparisons between \sys and 
collaborative fuzzing.
The most noticeable difference is that
\sys utilizes the \emph{runtime} information
in the fuzzer selection and resource adjustment, %
which results in subsequent differences.

\label{s:background}
\PP{Seed synchronization}
Seed synchronization~\cite{clusterfuzz-google-web,afl-zalewski-web, liang18:pafl} allows
the sharing of interesting seeds 
generated by different instances 
of the same fuzzer. 
Moreover, it helps
various modern fuzzers 
utilize multi-core processors~\cite{afl-zalewski-web,cgf-bohme-ccs16,lyu19:mopt,
lemieux18:fairfuzz,learnafl,chen2018angora,aschermann19:redqueen}.
It has been extended by~\cite{chen19:enfuz, osterlund21:collab}
to share unique test cases produced by different fuzzers.
In essence, 
seed synchronization introduces interesting inputs borrowed from other fuzzers
to a particular fuzzer that consistently fails to make progress.

\PN{\afl bitmap.}
Various numbers of fuzzers adopt \afl bitmap
to measure the performance of 
their fuzzing techniques~\cite{afl-zalewski-web, cgf-bohme-ccs16, bohme19:cover-based-greyb-fuzzin-markov-chain, lyu19:mopt, lemieux18:fairfuzz, learnafl, lafintel}. 
Recently, \angora\cite{chen2018angora} and AFL-Sensitive \cite{wang19:analyze-coverage} adopted
\afl bitmap in addition to call contexts
to implement a context-sensitive bitmap.
In essence,
\afl bitmap records the 
path explored during a fuzzing campaign,
measuring code coverage \cite{afl-detail}.
Also, the bitmap can be utilized 
as feedback to generate 
next round inputs.

\section{Design}
\label{s:design}
\algblock{Input}{EndInput}
\algnotext{EndInput}
\algblock{Output}{EndOutput}
\algnotext{EndOutput}
\newcommand{\Desc}[2]{\State \makebox[2em][l]{#1}#2}

\noindent We explain the design of \sys
to realize a meta-fuzzer
that allows
fine-grained and non-intrusive fuzzer selection.

\subsection{Overview of \sysn}
\label{ss:overview}
\sys aims to solve the \emph{selection problem}.
The key idea is
to dynamically deploy different sets of fuzzers
per workload
based on the fuzzer evaluation at runtime.
To this end,
\sys is composed of
two important components,
preparation phase (\autoref{ss:design:prep-phase}) and
focus phase  (\autoref{ss:design:focus-phase}).
In essence,
the preparation phase periodically monitors
the progress of individual fuzzers at runtime, called a trend,
and utilizes it as feedback
in its decision to select the next set of fuzzers.
Since the workload changes during fuzzing (\autoref{ss:eval-decision}),
the preparation phase helps \sys
adapt to the changed trend.
Thanks to the preparation phase,
the focus phase can prioritize the selected fuzzers
and achieve an optimal result
for any target binaries.

\begin{figure}[t]
  \begin{center}
    \includegraphics[width=\columnwidth]{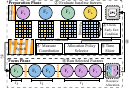}
  \end{center}
  \caption{
  Architecture overview of \sys. $F_1$ to $F_n$ are the baseline fuzzers.
	$bitmap_{1-n}$ captures the trend of the baseline fuzzers.
	The darker yellow colored
	region indicates that more paths have been discovered, which means strong trends.
	Based on the captured trends,
	\sys allocates resources to the selected fuzzers ($F_1, F_2, F_3$).
	For example,
	$F_1$ is allocated more resources compared with $F_3$,
	presented in time-sliced window $T_1$ and $T_3$ respectively.
	}
  \label{f:autofuzz-workflow}
\end{figure}

An overview of the two-phase design is presented in
\autoref{f:autofuzz-workflow}.
For a fair trend comparison among the baseline fuzzers,
\sys synchronizes the seeds of all fuzzers
in every round of the preparation phase (\WC{1}).
Also, it assigns the same amount of resources
to all baseline fuzzers.
Then,
each fuzzer takes its turn to run
in a very short time interval
until it encounters exit conditions (\WC{2},\WC{3}).
The details of the exit conditions are described in \autoref{ss:design:prep-phase}.
Then, \sys utilizes the evaluation result of the baseline fuzzers,
AFL bitmap,
to measure the trend of all fuzzers.
Considering the trend,
\sys selects a subset of the baseline fuzzers
along with the resource allocation meta-data,
deciding how each fuzzer should be prioritized
against the current workload (\WC{4}).
Note that this projection
exploits the fact that
the depicted trend will be highly maintained
during the focus phase (see \autoref{ss:eval-decision}
for detailed analysis).

Based on the resource allocation data,
\sys time-slices allocated CPU core(s) (\WC{5}).
\sys can support single-core and multi-core modes.
Single-core mode allows \sys to be integrated into the fuzzing benchmarks.
As a consequence,
\sys can take advantage of
all fuzzers supported by the FTS and \unifuzz.
Moreover,
\sys supports multi-core implementation (\autoref{s:impl}),
and a different number of cores can be assigned to each fuzzer
based on the resource allocation meta-data.
Even though the preparation phase is designed
to evaluate baseline fuzzers,
note that it can make some progress
because it actually runs the fuzzers.
Therefore,
seed synchronization before a transition to the focus phase
(\WC{6}, \WC{7})
allows the selected fuzzers to share
unique test cases produced during the preparation phase.

Then, the focus phase runs the selected fuzzers
one by one following the resource allocation meta-data.
Each fuzzer is allocated a specific CPU time window
to make progress (\WC{8}).
In the focus phase,
the goal is to achieve maximum performance,
not a fair comparison.
Therefore, after one fuzzer executes,
we synchronize the seeds
to allow the remaining fuzzers to explore
undiscovered paths by other fuzzers.
When the entire allocated resource is consumed,
it goes back to the preparation phase and
measures the trend (\WC{9}).
This flow of execution between the two phases
continues until the
fuzzing execution terminates (e.g., 24 hours).
A formal definition of the two-phase algorithm
can be found in \autoref{a:two-phase-combined}.

\PP{AFL bitmap as a unified metric}
Baseline fuzzers in \sys internally use
their original algorithms and metrics
(e.g., context-sensitive coverage of Angora,
block coverage of libFuzzer)
to evaluate progress and memorize interesting seeds and inputs
during both phases.
However,
fairly measuring their trends is difficult
when the metrics are incompatible;
it is difficult to tell that Angora outperforms libFuzzer
by comparing the context-sensitive coverage with block coverage.
Therefore, \sys adopts AFL bitmap as a unified criterion
to compare the progress of individual fuzzers (i.e., trends).
In detail, \sys runs the AFL-instrumented target
with the interesting input
found by each individual fuzzer's internal algorithm
to retrieve AFL bitmap of all baseline fuzzers
during preparation phases
(refer to \autoref{s:impl} for detailed implementation).

\subsection{Preparation Phase}
\label{ss:design:prep-phase}
\begin{algorithm}[h!]
  \caption{Preparation Phase}
  \label{a:prep-phase-with-early}
  \small
  \begin{algorithmic}[1]
    \Output
	  \State $Exit_{early} \gets$ Did preparation phase exit early?
	  \State $T_{remain}$ $\gets$ Remaining time of preparation phase
    \EndOutput
	  \Function{dynamic\_prep\_phase}{$\mathbb{F}$, $\mathbb{B}$, $T_{prep}$, $\theta_{cur}$, $C$}
	  \State $T_{remain} \gets T_{prep}$
	  \While {$T_{remain} > 0$}
	  \State $T_{run} \gets min(T_{remain},30)$
        \If {$C == 1$}
	   \For {\textbf{each} $f$ $\in$ $\mathbb{F}$}
	   \State \Call{run\_fuzzer}{$f$, $T_{run}$}
         \EndFor
        \Else \Comment{multi-core implementation}
         \State \Call{run\_fuzzers\_parallel}{$\mathbb{F}$, $T_{run}$, $\frac{C}{|\mathbb{F}|}$}
         \label{lst:prep-multi-core}
        \EndIf
	  \State $T_{remain} \gets T_{remain} - T_{run}$
	  \State $\pdiff \gets$ $\Call{find_best}{\mathbb{B}}$ - $\Call{find_worst}{\mathbb{B}}$
	  \If{$\pdiff > \theta_{cur}$}
	  \State \Return $(Exit_{early}=True$, $T_{remain})$
        \EndIf
      \EndWhile
	  \State \Return $(Exit_{early}=False, T_{remain})$
    \EndFunction
  \end{algorithmic}
\end{algorithm}
The goal of the preparation phase is
to appropriately select the fuzzers
that illustrate strong trends
to help the focus phase achieve maximum performance.
We describe
how the preparation phase measures the trends of baseline fuzzers
and automatically selects a subset
based on the trends.
We also introduce our novel approach,
which tries to minimize the waste of resources
caused by unfruitful runtime evaluation.
Last, we explain
how to efficiently distribute resources among the fuzzers
based on our novel strategy,
which can take advantage of both
individual fuzzers and collaborative fuzzing.

\PP{Dynamic time in preparation phase}
To adapt to the changed workload
and locate the best suitable fuzzers,
the preparation phase should evaluate all baseline fuzzers
until strong trends can be captured.
However,
if the time spent for the preparation phase is too long,
it can waste valuable resources
for the measurement.
Although \sys can also leverage the preparation phase
to make progress,
it can achieve better performance
by prioritizing the selected fuzzers earlier and longer.
On the contrary,
if the preparation time is too short,
capturing explicit trends of the fuzzers is difficult.
This makes \sys inappropriately prioritize
a suboptimal set of fuzzers
by precipitously assigning valuable resources to them.
As a consequence,
the trends cannot be sustained or
might be defeated by other fuzzers
during the focus phase.
In summary,
although the time allocated for the preparation phase is essential
to achieve optimal performance,
there is no oracle foretelling
how long the preparation phase should persist.
Instead of introducing another manual effort
to determine the proper time budget,
\sys introduces dynamic preparation time,
inspired by an Additive-Increase/Multiplicative-Decrease (AIMD) algorithm~\cite{aimd}.

\PP{Early exit and threshold}
To enable the dynamic time in the preparation phase,
\sys allows the preparation phase to exit
before the assigned time budget is completely consumed.
In essence,
if \sys can locate the fuzzers
that have strong trends earlier,
the remaining resources can be delegated to the focus phase.
However, to avoid the drawback of reducing the preparation time,
a premature decision,
\sys requires a clear indicator of strong trends.
As described in \autoref{a:resource-assignment},
we utilize the peak difference of the bitmap ($\pdiff$),
the difference between
the best- and worst-performing fuzzer.
After the early exit condition is detected,
\sys immediately enters the focus phase.
Note that the remaining time resulting from an early exit
will be delegated to the focus phase
so that it can
execute the selected fuzzer longer and earlier.
We introduce a threshold, presented as $\theta$,
that allows the preparation phase to exit early
if the peak difference is greater than $\theta$.

\PP {AIMD inspired threshold adjustment}
The threshold can be initially configured by users ($\theta_{init}$).
However, \sys automatically adjusts the initial configuration
and locates the optimal threshold per target,
which eliminates another manual effort.
As shown in ~\autoref{a:two-phase-combined},
the threshold is calibrated at every round
after preparation phases.
In detail, if an early exit happens,
\sys increases the threshold ($\theta_{cur}$) by $\theta_{init}$.
Otherwise, it divides $\theta_{cur}$ by two.
The rationale behind this design is
that the progress of the fuzzing campaign decreases
as it continues in general.
In its early phases,
due to the legitimate seeds provided as its initial input,
the selected fuzzer(s) generally produces
a fair amount of progress.
Thus,
if the threshold is too small,
it will be easily passed, and \sys can
make a suboptimal decision.
On the contrary,
as the fuzzing campaign continues,
the progress generated by each fuzzer is easily saturated
and exploring new paths becomes difficult.
Thus,
$\pdiff$ is frequently less than the threshold $\theta$
even if the best fuzzer performs much better than the others.
We show that \sys can
automatically configures $\theta$
(\autoref{ss:prelimanry-experiments}).

\PP{Trend evaluation}
As described in \autoref{a:resource-assignment},
we utilize bitmap operations to measure
the trend of individual fuzzers.
To this end,
\sys measures the unique paths that each fuzzer
has explored during the preparation phase.
Owing to seed synchronization before the preparation phase,
unique entries in the bitmap of each individual fuzzer
can represent its own contribution.
We present the common paths that have been
explored by all individual fuzzers
during the preparation phase
as $b_\cap$
(Line \ref{lst:bitmap-intersection}).
It is subtracted from
the bitmaps of individual fuzzers
so that the contribution of each fuzzer
can be measured
based on the unique paths discovered in the preparation phase.

\PP{Resource assignment}
\begin{algorithm}[t!]
  \caption{Resource Assignment Algorithm}
  \label{a:resource-assignment}
  \small
  \begin{algorithmic}[1]
    \Output
	  \State $\mathbb{RA} \gets$ $\{{r_1},{r_2},...,r_{n}\}$, $r_{n}$ is resource assignment for $f_{n}$
    \EndOutput

	  \Function{resource\_allocator}{$\mathbb{F}$, $\mathbb{B}$, $Exit_{early}$}
	  \State $\mathbb{RA}[f] \gets 0$, $\forall_f \in \mathbb{F}$
	  \State $b_{\cap} \gets$ $\bigcap_{i=1}^{n}b_{i}$ \label{lst:bitmap-intersection}
	  \State $b_{f} \gets b_{f} - b{\cap}, \forall_f \in \mathbb{F}$ \label{lst:bitmap-unique}

    \If{$Exit_{early} = True$}
    \For {\textbf{each} $f$ $\in$ $F$} \label{lst:max-fuzzers-start}
	  \If{$\Call{count}{b_{f}} > max\_count$}
	   \State $max\_count \gets \Call{count}{b_{f}}$
           \State $max\_fuzzers \gets \{f\}$
       \ElsIf{$c = max\_count$}
           \State $max\_fuzzers \gets max\_fuzzers \cup \{f\}$
       \EndIf
    \EndFor \label{lst:max-fuzzers-end}

	  \State $\mathbb{RA}[f] \gets \frac{1}{|max\_fuzzers|}$, $\forall_f \in max\_fuzzers$

	  \ElsIf{$Exit_{early} = False$}
	      \State $\mathbb{RA}[f] \gets \frac{\Call{count}{b_f}}{\sum_{i=1}^{n} \Call{count}{b_{i}}}$, $\forall_f \in \mathbb{F}$
       \label{lst:ra-proportional}
          \EndIf

	  \State \Return $\mathbb{RA}$
    \EndFunction
  \end{algorithmic}
\end{algorithm}
If some fuzzers perform well,
more resources will be allocated to them temporarily.
Therefore, \sys can accelerate well-performing fuzzers
and relegate the others.
\sys introduces novel strategies
that allow efficient resource distribution
among the selected fuzzers.
Thanks to the diversity in resource distribution,
\sys can take advantage of both worlds of
individual fuzzers and collaborative fuzzing.
Currently, \sys supports two resource allocation policies.
The first one prioritizes the best,
allocating all resources to the top-ranked fuzzer(s).
This policy is activated
only when a fuzzer significantly outperforms
the rest of all the baseline fuzzers.
Therefore, the $Exit_{early}$ signal
is utilized to detect this condition.
As described in ~\autoref{a:resource-assignment},
when the early exit occurs,
it first locates the set of fuzzers ($max\_fuzzers$)
that discovered the most unique paths
(Line \ref{lst:max-fuzzers-start}-\ref{lst:max-fuzzers-end}).
Note that multiple fuzzers can be selected
when there is a tie, but most of the time,
a sole fuzzer is selected.
Therefore,
by assigning all resources to the
best-performing fuzzer,
\sys can exploit the benefits of an individual fuzzer.
The other policy is
to proportionally distribute resources
based on the trends of each fuzzer.
Note that each contribution of baseline fuzzers is evaluated
based on the unique paths that each fuzzer has discovered.
Therefore,
if multiple strong trends are captured
during the preparation phase,
this highly indicates that
the preparation phase found
multiple fuzzers that
are favorable to fuzz
different parts of the program.
In that case,
we proportionally distribute resources to the fuzzers
based on their contribution (Line \ref{lst:ra-proportional})
and achieve the benefits of collaborative fuzzing.
Note that, unlike previous works,
\sys can distribute resources among the selected fuzzers
and achieve better performance.

\PP{Putting them all together}
As described in ~\autoref{a:prep-phase-with-early},
every round of the preparation phase requires
baseline fuzzers ($\mathbb{F}$), their bitmaps ($\mathbb{B}$),
time budget assigned for the current round ($T_{prep}$), and
threshold to detect early exit events ($\theta_{cur}$).
As a result,
the preparation phase returns $Exit_{early}$ indicating
whether the early exit occurs and
a non-zero value of $T_{remain}$ when an early exit occurs.
Also, it returns the $\mathbb{RA}$ resource allocation meta-data.
Each fuzzer belonging to $\mathbb{F}$
takes a turn to run
in a very short time interval, 30 seconds or $T_{remain}$.
After evaluating all fuzzers in $\mathbb{F}$,
the preparation phase checks whether the peak difference is
greater than $\theta_{cur}$
and early exits if the condition is met.
However,
if the difference is still under the threshold $\theta_{cur}$,
the preparation phase runs
each fuzzer for another short time interval.
The preparation phase will repeat this process until
either observing a large coverage difference
or spending all the predefined time budget ($\prep$).
We found that the short time interval assigned per fuzzer
does not significantly change the performance of \sys.
Therefore, we heuristically determine 30 seconds as the short interval.
However, if the interval is too short,
it will incur unnecessary context switches between fuzzers.
For multi-core implementation (Line \ref{lst:prep-multi-core}),
we run all fuzzers in parallel and distribute CPU resources equally.
\cc{RUN\_FUZZERS\_PARALLEL}($\mathbb{F}$, $T_{}$, c) runs
all fuzzers $f \in \mathbb{F}$ for $T_{}$ seconds,
and each fuzzer is assigned with $c$ CPU cores.

\subsection{Focus Phase}
\label{ss:design:focus-phase}
\autoref{a:focus-phase} describes
how the focus phase of \sys runs the selected fuzzers
utilizing the information generated by the preparation phase
of the same round.
Note that the list of fuzzers ($\mathbb{F}$) and
resource allocation meta-data ($\mathbb{RA}$)
are passed to the focus phase.
Because the time budget of each fuzzer
allowed to be consumed in this round
is measured based on the $\mathbb{RA}$,
no resources will be assigned to the fuzzers
that were not selected in the preparation phase.
Also, the focus phase requires $T_{focus}$,
which varies every round
depending on when the preparation phase exits
in this round.
Note that when the early exit occurs in the preparation phase,
the remaining time budget will be assigned to the focus phase
to allow it to utilize the strong trends longer.
The focus phase first calculates the total time budget (Line \ref{lst:focus-total})
and sorts the fuzzers based on the $\mathbb{RA}$
to run the well-performing fuzzers first (Line \ref{lst:sort-fuzzers}).
After that, it computes the time budget for the baseline fuzzers (Line \ref{lst:time-budget})
and runs the fuzzer if the assigned budget is non-zero (Line \ref{lst:focus-run1}-\ref{lst:focus-run2}).
Note that for every execution of one fuzzer,
the focus phase synchronizes the seed and bitmap (Line \ref{lst:focus-seed-sync})
to accumulate the progress of all selected fuzzers.
When multiple fuzzers are selected in the preparation phase,
the seed synchronization allows each fuzzer to
discover unique paths that have not been explored by others.
For the multi-core version,
we first calculate how many cores $c$ should be allocated
to each fuzzer
based on resource assignment $\mathbb{RA}$
(Line \ref{lst:focus-multi-core-cal}),
and then run all fuzzers in parallel
based on the result (Line \ref{lst:focus-multi-core}).

\section{Implementation}
\label{s:impl}
\sys consists of 6.2K lines (4.8K for the main framework, 1.4K for fuzzer API) of Python3 code. %
We implement our system as a Docker instance,
including benchmarks and fuzzers,
which makes \sys more portable and
its evaluation reproducible.
To restrict the resource usage of selected fuzzers
according to the resource allocation,
we utilize \textit{cgroups} \cite{cgroups},
which can manage processes in hierarchical groups
and limit the resource usage per group.

\PN{\AFL bitmap measuring coverage of fuzzers.}\
In \sys,
each baseline fuzzer utilizes
a fuzzing algorithm of its original design,
which does not involve any implementation changes.
Therefore,
during the two phases,
each fuzzer can generate different interesting inputs
based on its internal metrics and algorithms.
However, as described in \autoref{ss:overview},
\sys needs to retrieve \afl bitmap of all baseline fuzzers
to fairly compare the trends.
Specifically,
\sys invokes the \afl-instrumented binary
with different inputs
found to be interesting by each individual fuzzer.
Each fuzzer can maintain
the interesting inputs as files
in different directories.
Therefore,
when the new fuzzer is integrated into \sys,
this information should be known to \sys.
For example,
\angora configures
"\cc{queue}", "\cc{crashes}", and "\cc{hangs}"
as interesting input directories.
Therefore,
\sys invokes the \AFL-instrumented target
with new input and measures \AFL bitmap coverage of \angora
whenever a new file is created
in one of the specified directories during the preparation phase.

\PP{API for integration.}
Any individual fuzzer implementing
the following python APIs
can be integrated into \sys.
\begin{enumerate*}
\item Start / Stop
\item Scale up / down (for parallel mode)
\end{enumerate*}.
Each fuzzer requires different arguments and file directories
to initiate the fuzzing.
The Start and Stop API makes \sys understand
how to initiate and stop each fuzzer
with a proper argument.
Also, each fuzzer should implement
the Scale Up and Down API
to take advantage of multi-core resources.
For example, \afl has master and slave modes.
Therefore, the user needs to implement the Scale Up API
to launch more slave instances
instead of master instances.

\PP{Multi-core support}\
\sys is able to utilize multiple cores.
For the multi-core implementation,
we concurrently run the fuzzers in the preparation phase and the focus phase.
For example,
if we have $N$ cores,
\sys instantiates $\lceil N \times \mathbb{RA}[f] \rceil$ processes
using the Scale Up API for fuzzer $f$.
After that,
\sys utilizes \textit{cgroups}
to limit the exact total CPU resources allocated
to the generated fuzzer instances as $N \times \mathbb{RA}[f]$.
\begin{figure*}[t]
  \begin{center}
    {
    \fontsize{6pt}{6pt}\selectfont
     \def\svgwidth{\textwidth}
     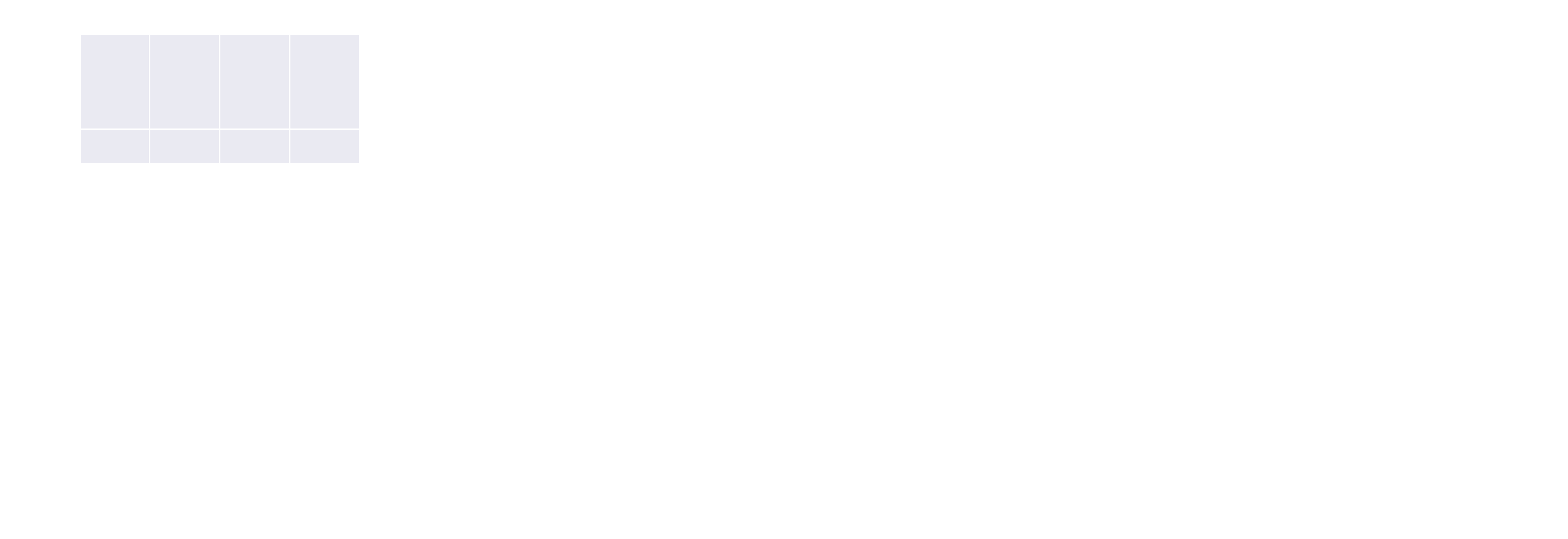
    }
  \end{center}
    \vspace{-0.4cm}
	\caption{Evaluation of \sys on \unifuzz and \fts.
	Each line plot is depicted with an arithmetic mean
	and 80\% confidence interval for 10 times fuzzing executions.
	Coverage ratio is the percentage of branches explored by each fuzzer.}
  \label{f:autofuzz-eval-best}
\end{figure*}
\section{Evaluation}
\label{s:eval}
\noindent We evaluate \sys by answering the following questions.
\begin{itemize}[noitemsep,nolistsep]
\setlength\itemsep{0em}
\item \PPI{Target independent fuzzer selection}
How can \sys effectively achieve
better code coverage against different target binaries
compared with individual fuzzers? (\autoref{ss:overall-vs-individual})
Does the runtime information
allow \sys to exceed collaborative fuzzing
with manual analysis? (\autoref{ss:autofuzz-enfuzz-cupid})
\item \PPI{Non-intrusive fuzzer selection.}
Does the initial configuration for
preparation and focus time
affect the efficiency of \sys? (\autoref{ss:prelimanry-experiments})
How do the early exit and AIMD
allow \sys to locate a suitable set of
preparation and focus time at runtime? (\autoref{eval:resource distribution})
\item \PPI{Dynamic Resource allocation strategy.}
How can \sys's resource allocation
prioritize the well-performing fuzzers
based on the trends
compared with previous works? (\autoref{ss:eval-num-fuzzers})
\item \PPI{Number of bugs found.}
Higher coverage does not necessarily lead to more bugs. Can \sys outperform other fuzzers in terms of bug finding? (\autoref{ss:eval-bug-discovery})
\item \PNI{Accuracy of decisions made by \sys.}
How accurate is the decision of \sys?
Does the resource allocation decision
allow \sys to achieve optimal results?
(\autoref{ss:eval-decision})
\end{itemize}

\onlymain{
Because of space limitation, a part of result is only available in the extended
  version \cite{autofz-extended}.
}

\subsection{Experimental Setup}
\label{ss:setup}

\PP{Host environment.}
We evaluated the following experiments
on Ubuntu 20.04 equipped with
AMD Ryzen 9 3900 having 24 cores and 32\GB memory.
To compare \sys with \cupid and \enfuzz,
we assign \emph{multiple CPU cores} to
a Docker container.
The rest of the evaluations
comparing \sys with individual fuzzers
are executed
with a container to which
\emph{one CPU core} is assigned without a memory limit.
All containers run Ubuntu 16.04\footnote{We chose Ubuntu 16.04 because it is the latest version that successfully builds all fuzzers and benchmarks used in the evaluation.} because of compatibility.

\PP{Baseline fuzzers.}
For the evaluation,
\sys employs all fuzzers supporting seed synchronization
from \unifuzz and \fts.
Also, \sys supports all the fuzzers
adopted in \cupid and \enfuzz
for a fair comparison. %
All the fuzzers supported by \sys are
\afl\cite{afl-zalewski-web},
\aflfast\cite{cgf-bohme-ccs16,
  bohme19:cover-based-greyb-fuzzin-markov-chain}, \mopt\cite{lyu19:mopt},
\fairfuzz\cite{lemieux18:fairfuzz},
\learnafl\cite{learnafl},
\qsym\cite{yun18:qsym},
\angora\cite{chen2018angora},
\redqueen\cite{aschermann19:redqueen},
\radamsa\cite{radamsa}, \lafintel\cite{lafintel}, and
\libfuzzer\cite{serebryany15:libfuzzer}.
We utilize the modified version of \libfuzzer
excerpted from \cite{guler20:cupid}
because it does not support seed synchronization. %
Also,
we use the implementation provided by~\cite{fioraldi20:afl++}
for \radamsa, \redqueen, and \lafintel.
Note that
\sys can adopt any fuzzer supporting seed synchronization,
and no fundamental obstacles
prevent their integration.
Therefore, \sys can truly enjoy the benefit of
increasing diversity of the fuzzers.

\PP{Target binaries and seeds}
We integrate \sys into
\unifuzz \cite{li21:unifuzz} and \fts \cite{fts}
so that we can evaluate \sys
on various real-world programs.
We excluded targets that cannot be compiled by all fuzzers and targets of which coverage saturates very early.
We believe that
diversities in the targets are enough to demonstrate
the versatility of \sys.
We adopt the default seeds
provided by the benchmarks.

\subsection{Comparison with Individual Fuzzers}
\label{ss:overall-vs-individual}
\begin{figure}[h]
  \centering
    \includegraphics[width=\columnwidth]%
    {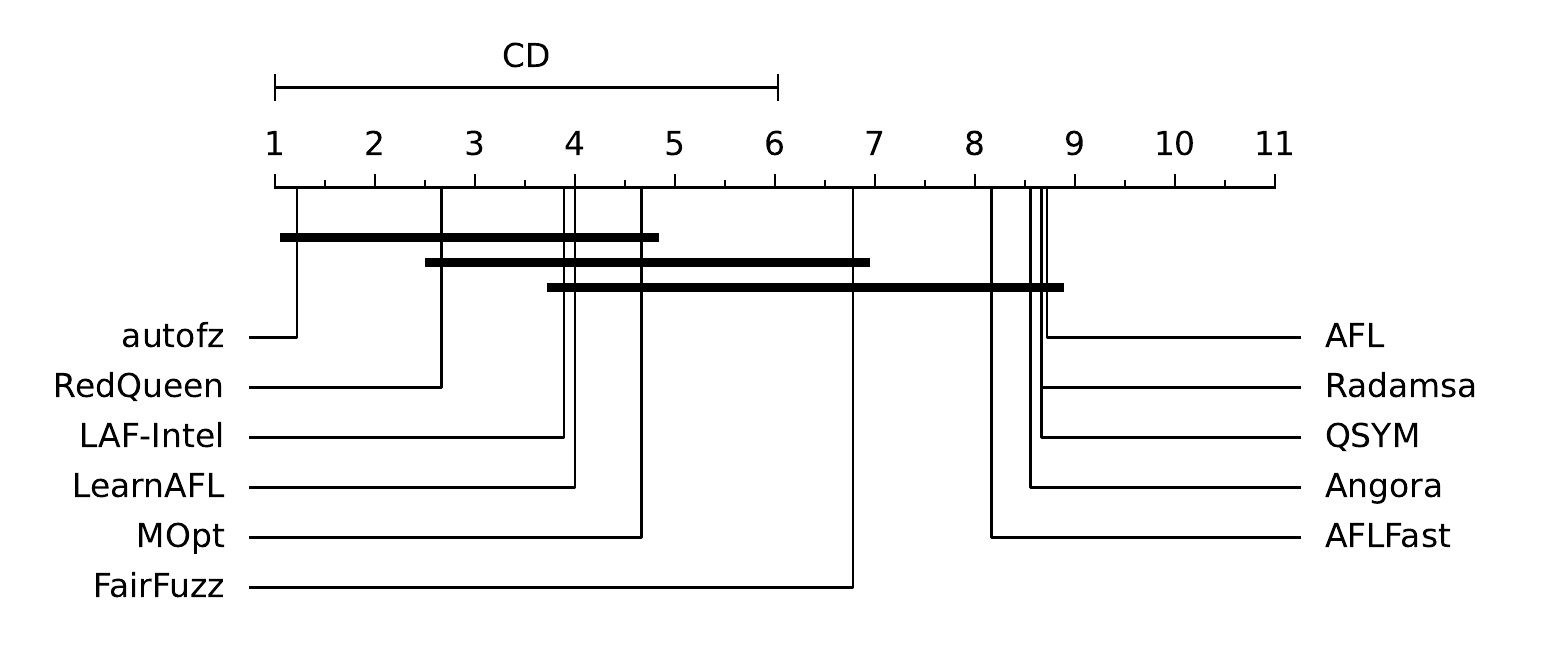}
  \vspace{-0.6cm}
  \caption{
    Critical Difference for the targets in \unifuzz.
    Each number
    indicates
    the average rank of fuzzers.
    Each bold line
    indicates that
    there is no statistical difference in performance
    among the fuzzers grouped by that line
    in terms of the Nemenyi post-hoc test.}
  \label{f:cd-unifuzz}
\end{figure}

To evaluate the efficacy of \sys,
we measure the AFL bitmap coverage
on \fts and \unifuzz benchmarks.
We configure all parameters described in~\autoref{a:two-phase-combined}
as follows: $\prep =300$, $\focus = 300$ and $\theta_{init} = 100$.
See \autoref{ss:prelimanry-experiments}
to understand how different parameters affect
the performance of \sys.
The coverage graphs
are depicted in \autoref{f:autofuzz-eval-best}.
\sys ranks best in almost all benchmarks and only loses to \redqueen on
\pdftotext, which proves that \sys outperforms
individual fuzzers.

To understand how frequently
\sys can outperform individual fuzzers
across various targets,
we introduce Critical Difference (CD) diagrams~\cite{demvsar:cd} used in~\cite{fuzzbench}.
\autoref{f:cd-unifuzz} and \autoref{f:cd-fts} depicts the critical difference
based on the averaged ranks of individual fuzzers and \sys
on \unifuzz and \fts, respectively.
On average,
\sys ranks 1.22 and 1.2 in \unifuzz and \fts, respectively.
This evaluation demonstrates that
the runtime trend allows
\sys to outperform
all individual fuzzers
regardless of the targets.
This result is important
because the user does not need to spend
valuable computation power to understand
which fuzzers are suitable for fuzzing
particular binaries.

In addition to the average ranks,
we report the detailed coverage of individual fuzzers and \sys
in \eautoref{t:cov-vs-focus}
to further highlight how significant the differences are.
Furthermore, we provide the Mann–Whitney U Test
\cite{klees18:evaluat-fuzz-testin, arcuri12:random-testing}
between \sys and other individual fuzzers one by one
in \eautoref{t:p-value-vs-focus}.
This evaluation demonstrates that
\sys is statistically differentiated from
most individual fuzzers (p-value $< 0.05$)
on overall benchmark suites.

\subsection{Elasticity of \sysn}
\label{ss:prelimanry-experiments}
\begin{figure}[h]
  \begin{center}
    \includegraphics[width=\columnwidth]{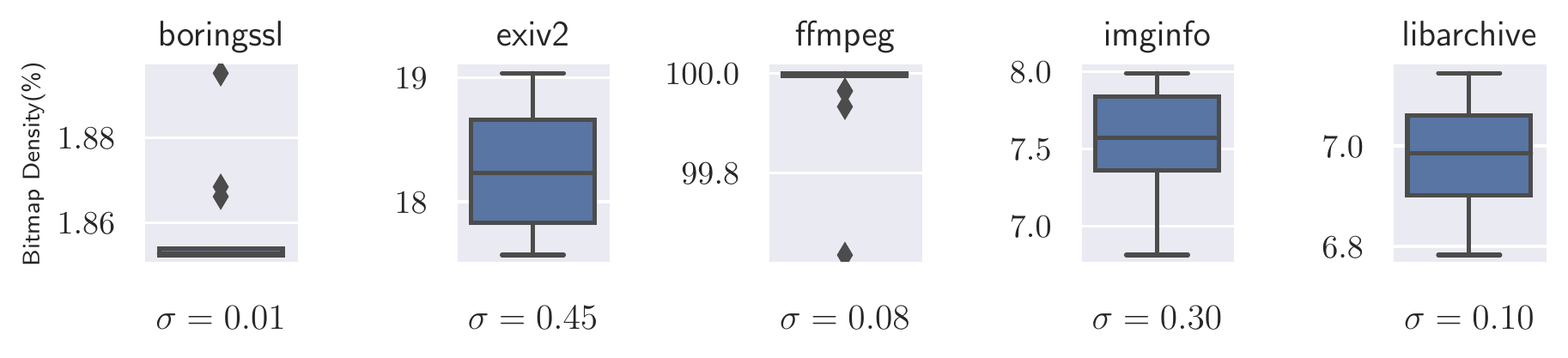}
  \end{center}
    \vspace{-0.45cm}
	\caption{
	    The graph shows
    the coverage distribution of all configurations. $\sigma$ is the standard deviation.
  }
  \label{f:cd-combined-aimd-parameter}
\end{figure}

As described in~\autoref{a:two-phase-combined},
\sys requires the user to configure three parameters:
$T_{prep}$, $T_{focus}$, and $\theta_{init}$.
With the initial configuration,
\sys automatically locates subsequent parameters
such as a suitable set of fuzzers,
corresponding resource allocation,
and the proper time to switch the fuzzers.
In this evaluation,
we argue that the design of \sys is elastic enough
to achieve good performance in general
\emph{independent of} the initial configurations.
Our evaluation includes multiple combinations of
three variables as specified in the following.
\begin{itemize}[noitemsep,nolistsep]
\item $T_{prep}$: 300; $T_{focus}$: 300, 600, 900
\item $\theta_{init}$: 10, 100, 200, 300, 400, 500
\end{itemize}
We compare the average rank of
18 different configurations of \sys
(enumerated in \eautoref{f:cd-combined-aimd-parameter-appendix})
for the five targets presented
in~\autoref{f:cd-combined-aimd-parameter}.
The evaluation demonstrates that
the different configurations do not
result in noticeable contrast in performance.
Note that the standard deviation, $\sigma$,
is very small in all targets,
indicating
there is a very small numerical difference
in bitmap-wise comparison.
Moreover,
we randomly select a subset of benchmark targets
to further prove that
\sys performs well on any targets,
independent of the configurations.
As shown in~\autoref{f:autofuzz-eval-best},
we deploy \sys with the best configuration found in
\eautoref{f:cd-combined-aimd-parameter-appendix}
($\prep = 300$, $\focus = 300$, $\theta_{init} = 100$)
and observed that the selected configuration allows \sys to perform well
for most of the benchmark targets
that have not been selected in evaluating this configuration.

\subsection{Resource Distribution in Actions}
\label{eval:resource distribution}

\begin{figure}[h!]

\begin{subfigure}{\columnwidth}
    {\rowcolors{2}{}{gray!20}
  \begin{center}
    \begin{tabular}{lrrrrr}
\hline
Round & Winner & $\pdiff$ & $\theta$ & $\prep$ & $\focus$  \\ \hline
   $1^*$ & \angora   & 857 & 100 & 30 & 570 \\ \hline
   $2^*$ & \redqueen & 234 & 200 & 150 & 450 \\ \hline
   3 & None      & 116 & 300 & 300 & 300 \\ \hline
\end{tabular}
  \end{center}
  \vspace{-0.5em}
  \footnotesize{* asterisk mark after the round numbers means that $Exit_{early}$ is true.}

}
\label{t:exiv2-round-winner}
\end{subfigure}

\begin{subfigure}{\columnwidth}
  \centerline{
     \includegraphics[width=\columnwidth]{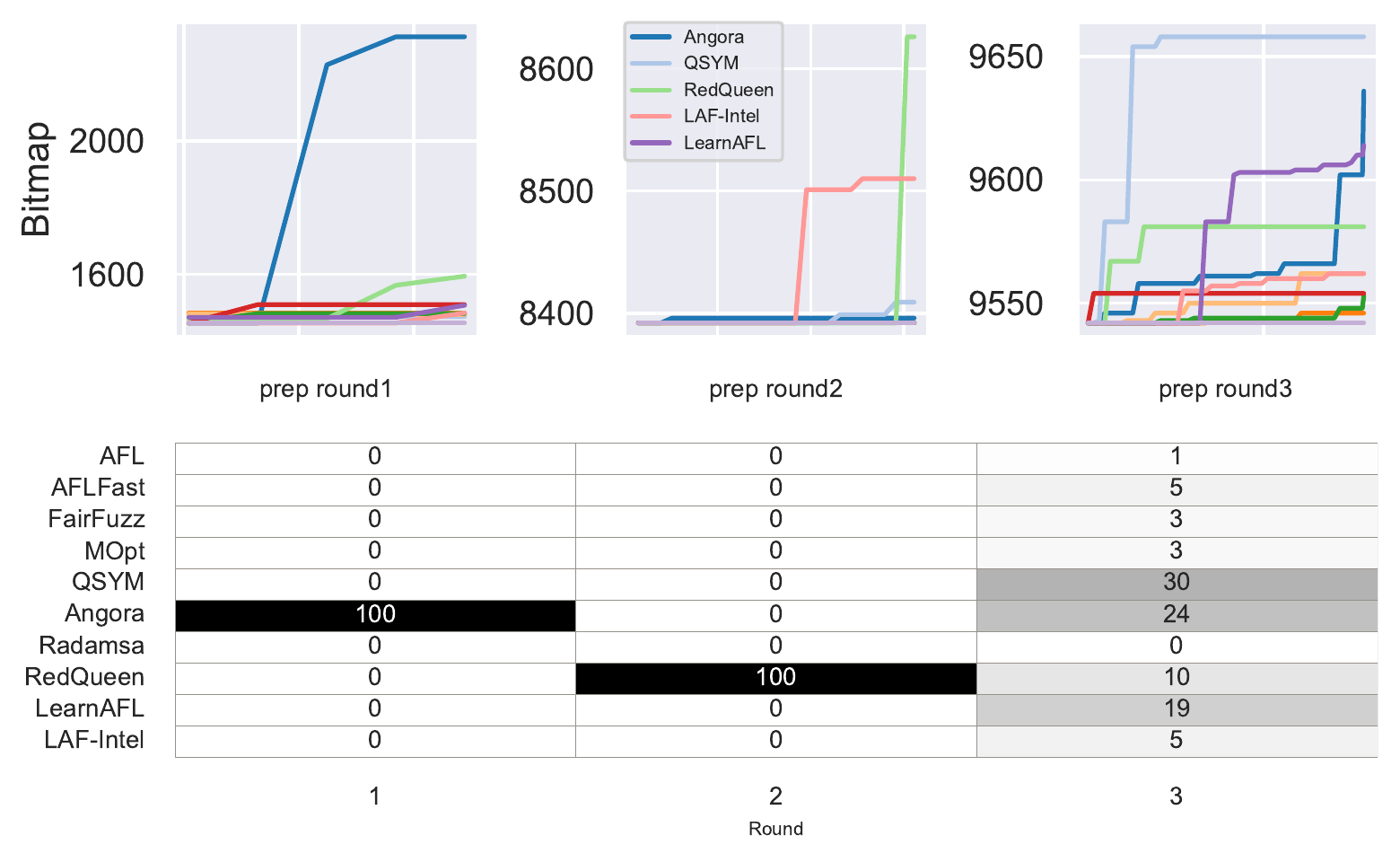}
  }

\end{subfigure}

\caption{
Resource distribution decision of \sys for \exiv2.
The first table shows
how the parameters
associated with the early exit change
during the first five rounds of \sys.
$T_{prep}$ indicates a preparation time \emph{actually spent}
in a particular preparation round,
not the time initially allotted
to a preparation phase.
The graphs in the second row depict
captured trends of some baseline fuzzers
during the preparation phases. Note that it does not start with $y=0$ to highlight the difference between fuzzers.
The heat map in the last row shows
resource allocation decision
made by each preparation round.
Refer to
\autoref{t:exiv2-round-winner-full}
for remaining rounds.
}

\label{f:resource-distribute-exiv2}
\end{figure}

In this section,
we demonstrate
how \sys can transform
runtime trends of different baseline fuzzers
into a resource allocation decision.
Furthermore, we explain
how the early exit events (\autoref{ss:design:prep-phase}),
together with the resource allocation,
helps \sys take advantage of the strong trend of individual fuzzers
while maximizing the collaboration
effects of multiple different fuzzers.
We depict the first three rounds of evaluation of \sys
to highlight different aspects of resource allocation
on \exiv2 target.

In the first preparation phase,
most of the individual fuzzers
can discover a fair amount of unique paths
thanks to the provided initial seeds.
However, \angora significantly outperforms others.
As shown in \autoref{f:resource-distribute-exiv2},
the first preparation phase exits very early ($T_{prep}, 30$)
because it found that
the $\pdiff$, measured as 857,
exceeds the initial threshold value ($\theta$, 100).
Therefore,
\sys allocates all resources to \angora
following the \autoref{a:resource-assignment}
in the hope that the strong trend of \angora would
continue during this focus phase.
Because the preparation phase exits early,
the remaining time is assigned to the focus phase ($T_{focus}, 570$).
In the second preparation round,
as the $\theta$ is increased to 200
due to early exit in the first round
(following \autoref{a:two-phase-combined}),
\sys did not exit early
even though \lafintel started to perform well
compared to the others ($\pdiff$, $\approxeq 100$).
Therefore,
\sys continued the preparation phase
until ($T_{prep}, 150$) the best (\redqueen) surpassed the lowest
by the $\theta$, 200
(see the graph for prep round2).
This result demonstrates that
\sys prevents it from choosing
the suboptimal set of fuzzers.
\begin{figure*}[t!]
  \begin{center}
    \includegraphics[width=\textwidth]{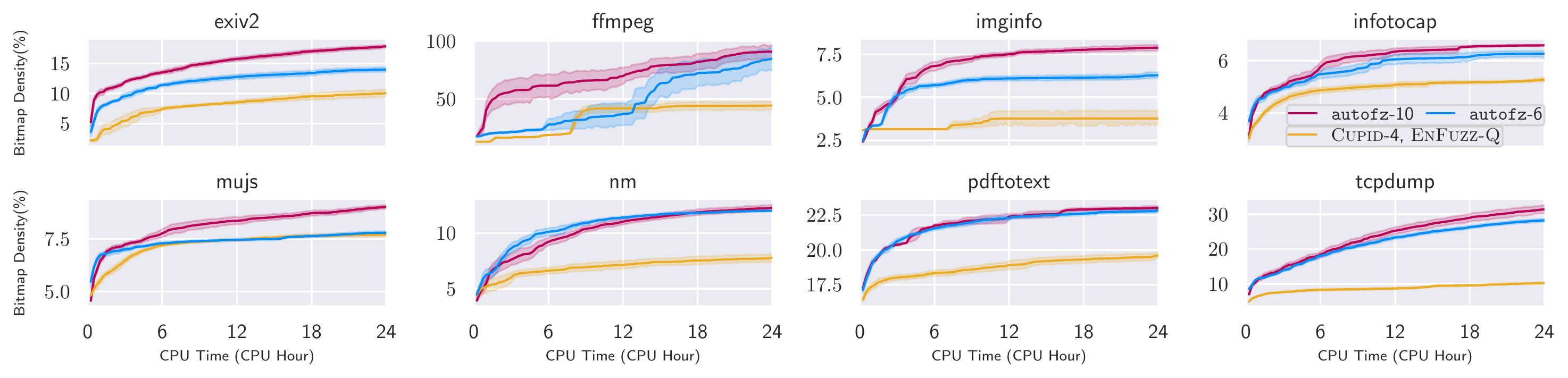}
  \end{center}
  \vspace{-0.3cm}
	\caption{
          Comparison among \sys, \enfuzz, and \cupid on \unifuzz.
          Each line plot is depicted with an arithmetic mean
          and 80\% confidence interval for 10 times fuzzing runs.
          The coverage ratio is the percentage of branches explored by each fuzzer.
          X-axis is elapsed \textbf{\textit{CPU time}}.
          Below is the list of fuzzers used by different configurations:
          \textbf{\sys-6} = [\afl, \fairfuzz, \qsym, \aflfast, \lafintel, \radamsa],
          \textbf{\cupid-4 and \enfuzz-Q} = [\afl, \fairfuzz, \qsym, \aflfast],
          \textbf{\sys-10} = [All baseline fuzzers described in~\autoref{s:eval} except \libfuzzer]}
  \label{f:autofuzz-enfuzz-cupid-unifuzz}
\end{figure*}
In the third preparation phase,
\qsym exhibits reasonably good performance compared to the others ($\pdiff, 116$).
However,
most of the baseline fuzzers achieve similar coverage
during the preparation phase (see the graph for prep round 3),
and it cannot exceed the threshold ($\theta, 300$)
to exit early in the third preparation phase.
Therefore,
\sys decides to allocate the resources proportionally
based on the trends of each individual fuzzer,
unlike the first two rounds
dedicating all resources to a sole fuzzer.

\PN{Slow but gradually improving fuzzers.}
Depending on the complexity of the baseline fuzzers,
the time assigned for preparation phases
might not be sufficient
to build internal metrics.
For example,
\qsym requires a long initialization time
for its concolic executor to build up internal states.
Moreover,
the concolic executor is designed to solve
hard branches, so it might not be efficient to
explore easy branches,
which causes the other lightweight fuzzers
to be selected until easy branches are resolved.
As a result,
fuzzers like \qsym can be treated as
relatively underperforming, especially
in the first few rounds or
when the early exit occurs frequently,
which causes it not to be selected in the focus phase
and further prevents it from building up
an internal state.
However, thanks to the dynamic nature of \sys,
it will allocate more time resources
for preparation phases
when other fuzzers become saturated
(i.e., resolving all easy branches)
and naturally take care of such
slow but gradually improving fuzzers.
As shown in \autoref{t:exiv2-round-winner-full},
\qsym has not been selected in most of the early rounds,
but all resources are assigned to it
in later rounds (i.e., 12 and 15).

\subsection{Comparison with Collaborative Fuzzing}
\label{ss:autofuzz-enfuzz-cupid}
In this section,
we compare \sys with collaborative fuzzing
such as \enfuzz and \cupid
for \unifuzz and \fts targets.
We describe the result of \unifuzz
to demonstrate that
the prior knowledge utilized in collaborative fuzzing
can be a major hindrance
to efficient and automatic fuzzer selection.
We also present the comparison
for \fts targets in \autoref{s:fts-collab}.

\PP{CPU hours.}
We utilize the \textbf{multi-core} implementation of \sys
because collaborative fuzzing
concurrently deploys multiple fuzzers.
To further achieve fairness in comparison,
we fixed the total CPU time to 24 hours.
The CPU time represents the total time resources
consumed by all cores in each configuration.
For example, \sys-6 and \cupid utilize six and four cores each;
if we execute both configurations in \emph{physical 24 hours},
each configuration spends 6*24 and 4*24 CPU hours, respectively,
which is unfair to \cupid.
Therefore, we make all evaluations run \emph{24 CPU hours} in total.
If one configuration utilizes $N$ cores,
each core can run $\frac{24}{N}$ hours.

\PN{Fuzzer selections for \unifuzz}
Since the evaluations and fuzzer selection
of \cupid and \enfuzz
target \fts,
we cannot directly adopt
the set of fuzzers depicted as the best
in their works.
Note that their best configurations
include \libfuzzer,
and \unifuzz does not support it.
Therefore,
instead of \libfuzzer,
we selected \lafintel for \sys-6 and \cupid-4
in addition to five fuzzers utilized
as the baseline in both works.
For \cupid,
we additionally retrieve its artifact and select
the best four-fuzzer combination
reported by the artifact.
The detailed list of fuzzers and evaluation results is illustrated in~\autoref{f:autofuzz-enfuzz-cupid-unifuzz}.
We used the same set of fuzzers
depicted as the best in their works
in comparison for \fts targets (refer to ~\autoref{f:autofuzz-enfuzz-cupid}) because \fts supports \libfuzzer.

\PP{No prior knowledge, but a promising result.}
The set of fuzzers selected by \cupid-4
is derived from the baseline fuzzers
used by \sys-6,
relying on offline analysis.
However, \sys-6 outperforms \cupid-4 by 83.63\%
without any prior knowledge
(see \eautoref{t:cov-autofuzz-cupid-enfuzz-unifuzz}).
Considering that \sys-6 and \cupid-4
achieves similar result for \fts targets
(see \autoref{f:autofuzz-enfuzz-cupid}),
we can infer that
prior knowledge used in fuzzer selection
is not diverse enough
to represent real-world binaries.
Note that the training set used by \cupid is generated
as a result of running a subset of \fts targets;
thus it can be biased toward \fts
and cannot reduce the chance of overfitting
when fuzzing \unifuzz targets.
Our evaluation empirically demonstrates that,
when relying on prior knowledge in fuzzer selection,
results can be suboptimal and
cannot guarantee the best outcome
independent of the targets.
Moreover,
\sys makes evident that
runtime trends are more reliable and cost-effective
than
well-established prior knowledge
in fuzzer selection.
Also, note that
the selected set of fuzzers for \cupid-4
is the same as \enfuzz-Q,
one of the configurations
handpicked by
the \enfuzz authors.
This further indicates that
expert-guided, target-specific
combinations are unreliable
and do not yield the best outcomes.

\begin{figure}[t]
  \centering
  \includegraphics[width=\columnwidth]{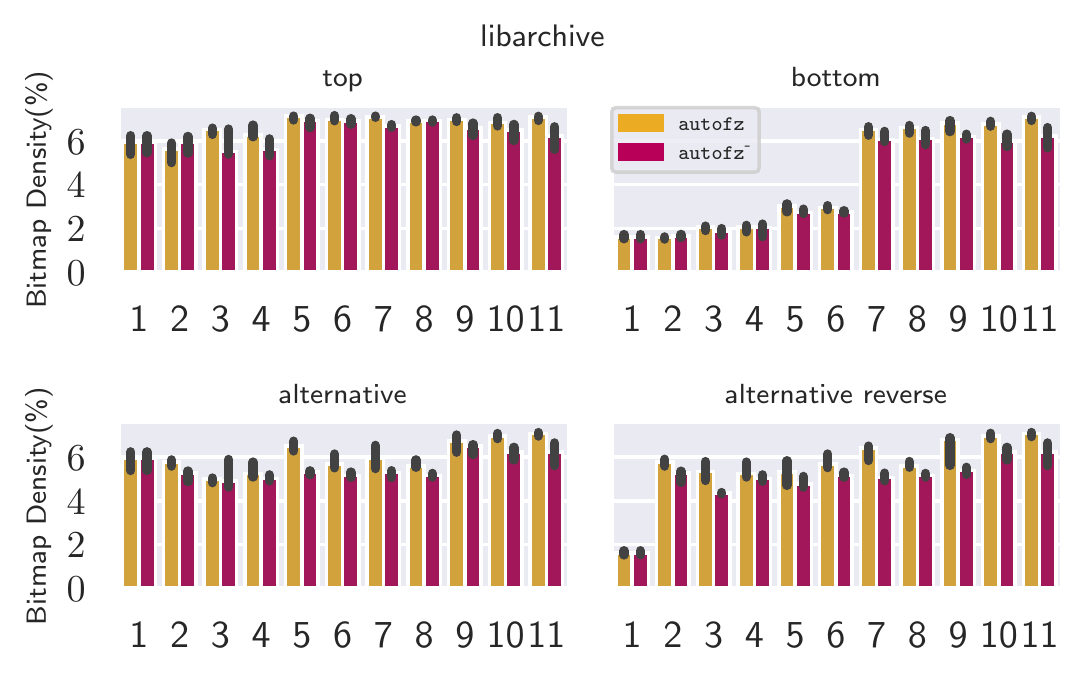}
  \caption{
  Evaluation of \sys and \sysm on
  different numbers of baseline fuzzers. The target is \cc{\libarchive}.
  The X-axis indicates the number of fuzzers included to the baseline.
  The bar plots represent the average bitmap density across 10 executions.
  The accompanied error bars are generated with an 80\% confidence interval.
  At $x=1$, \sys and \sysm are identical. It is included
  as a baseline to be compared with different configurations.
  }
  \label{f:fuzzer-num-libarchive}
\end{figure}

\subsection{Effects of the number of baseline fuzzers}
\label{ss:eval-num-fuzzers}
As shown in \autoref{f:autofuzz-enfuzz-cupid} and \autoref{f:autofuzz-enfuzz-cupid-unifuzz},
both \sys-11 and \sys-10 outperform \sys-6
in all benchmarks suites, except \boringssl.
This implies that
adding more fuzzers
will achieve better coverage in general
due to the different natures induced by various fuzzers.
However, the relationship between
the number of fuzzers and their performance is not straightforward.
This relationship can also be affected by
how effective the resource scheduling algorithm of \sys is.
Therefore, we introduce an \sysm, a variant of \sys,
which equally allocates resources to all baseline fuzzers
in round-robin fashion.
In this section,
we evaluate \sys and \sysm
with various numbers of baseline fuzzers
to provide a profound understanding of
how this number can affect their performance
in accordance with the underlying resource allocation algorithm.
Also,
we provide two thorough case studies.

\PP{Evaluation compositions.}
In evaluating delta
induced by having another baseline fuzzer,
what fuzzer will be included to the baseline is important.
For example,
introducing a well-performing fuzzer
gives a high chance of performance increase,
but adding a bad fuzzer will not guarantee
an improvement as much as the good one.
Therefore,
we test \sys and \sysm
with four different sequences in fuzzer addition:
\emph{top}, \emph{bottom}, \emph{alternative}, and \emph{alternative reverse}.
We explain the differences of these four different sequences
using \libarchive as an example.
We evaluate each fuzzer's performance
based on \autoref{f:autofuzz-eval-best}.
The \emph{top} sequence introduces fuzzers
in descending order of performance: \redqueen,  \lafintel, \libfuzzer, \qsym,
\angora, \mopt, \learnafl, \fairfuzz, \afl, \aflfast, and \radamsa.
\ignore{On the contrary} Conversely, the \emph{bottom} brings in
new fuzzers in ascending order (from the least coverage).
For \emph{alternative},
it selects one in the best and the next one in the worst, alternatively.
For example, %
it selects \redqueen (\nth{1}), \radamsa (\nth{11}), \lafintel (\nth{2}), \aflfast
(\nth{10}) $\cdots$.
The \emph{alternative reverse},
selects the one in the worst and the next one in the best one by one.
Therefore, the order will be  \radamsa (\nth{11}), \redqueen (\nth{1}), \aflfast (\nth{10}),
\lafintel (\nth{2}) $\cdots$.

\PN{\sys versus \sysm.}
The comparison between \sys and \sysm,
described in~\autoref{f:fuzzer-num-libarchive},
highlights the importance of resource allocation.
It demonstrates that
\sys can outperform \sysm in most of the cases
if both utilize the same fuzzer set.
With the \emph{top} and \emph{bottom} cases,
we see that
the performance difference between \sys and
\sysm is not very large initially but
increases as more fuzzers are introduced.
The performance gap between \sys and \sysm is much larger
in the \emph{alternative} and \emph{alternative reverse} cases
in almost all cases.
Remember that \sysm equally distributes resources
to all baseline fuzzers,
but \sys can redistribute resources
based on the trends of individual fuzzers
per workload.
Therefore,
fewer resources will be assigned
to well-performing fuzzers
as more fuzzers are introduced in \sysm.
\emph{This shows that
\sys can prioritize well-performing fuzzers
and relegate poorly performing fuzzers automatically}
on account of its clever resource allocation algorithm.
We present further evaluations on \nm and \exiv2
in \eautoref{f:fuzzer-num-exiv2}.

\begin{figure}[tb]
  \centering
  \begin{subfigure}[b]{\columnwidth}
    \includegraphics[width=\columnwidth]{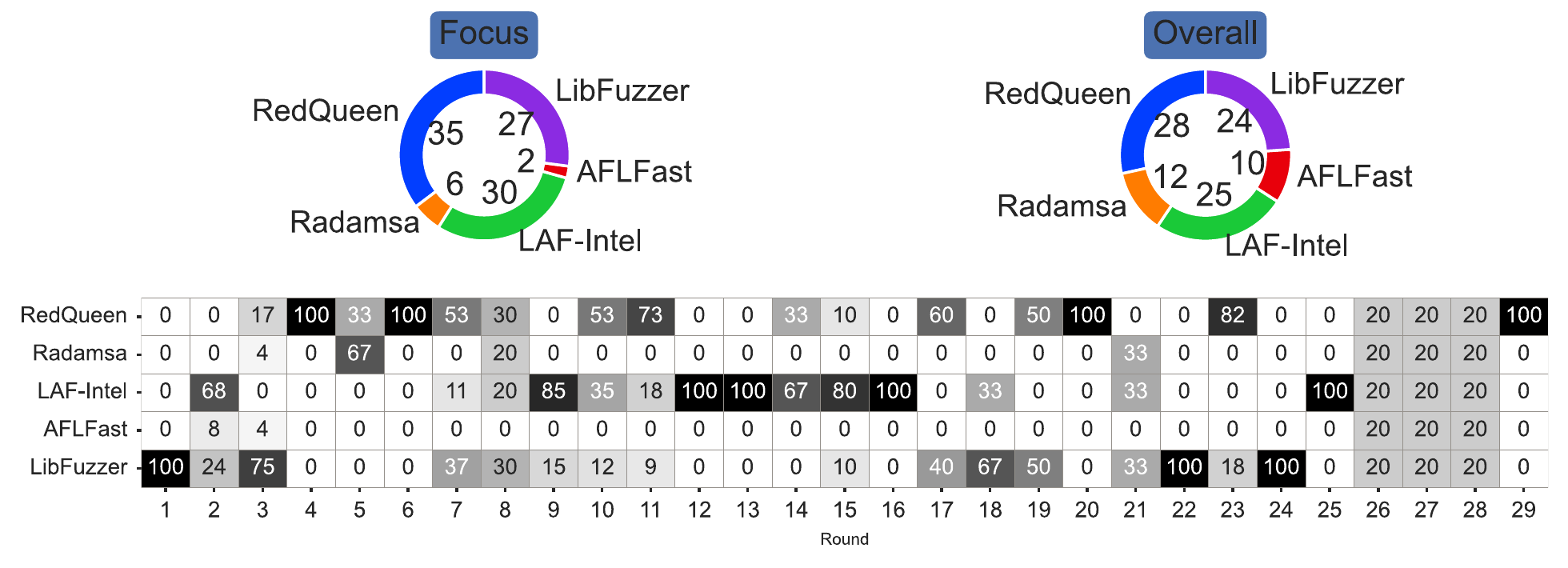}
    \caption{5-fuzzers (\redqueen, \radamsa, \lafintel, \aflfast, \libfuzzer)}
    \label{f:autofuzz-libarchive-alt5}
  \end{subfigure}
  \begin{subfigure}[b]{\columnwidth}
    \includegraphics[width=\columnwidth]{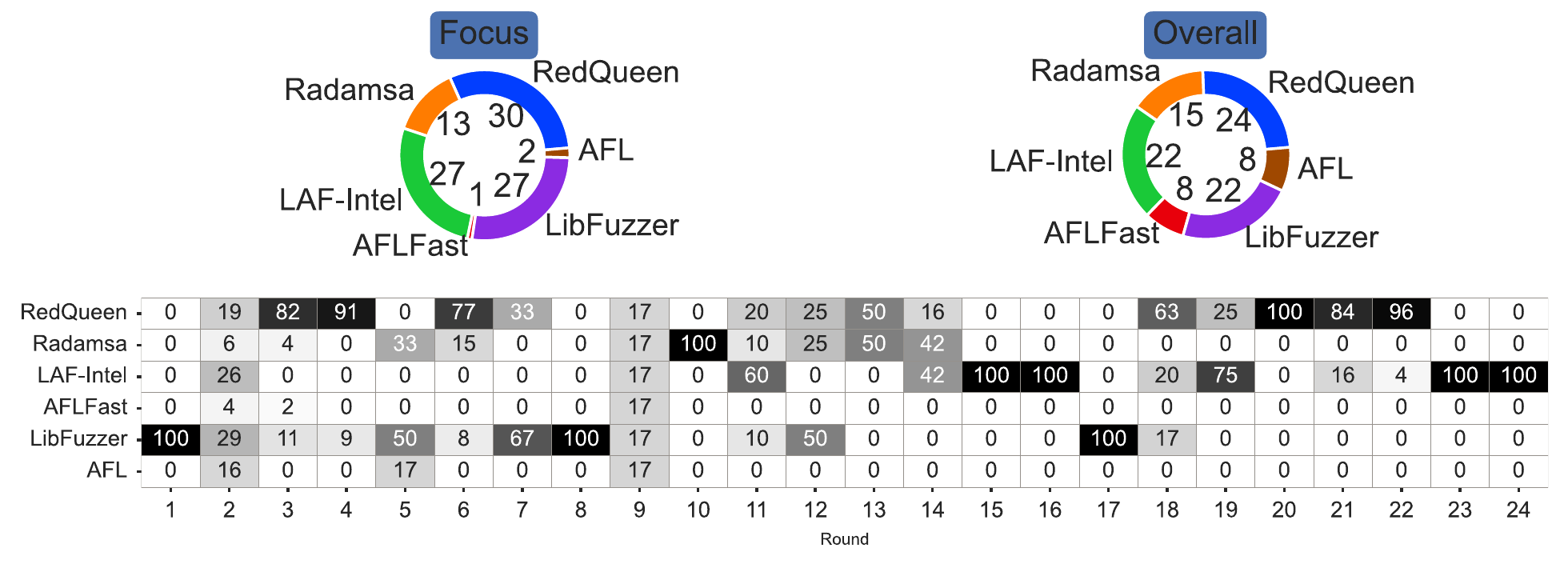}
    \caption{6-fuzzers (5 fuzzers in~\autoref{f:autofuzz-libarchive-alt5} plus \afl)}
    \label{f:autofuzz-libarchive-alt6}
  \end{subfigure}
  \caption{
  Each pie chart presents the
  aggregated results across all rounds of the focus phases
  and overall fuzzing campaign (including preparation phases),
  respectively.
  The heat map describes the resource allocation
  determined by every round of the preparation phase.
  All figures are populated with data
  presented in \autoref{f:fuzzer-num-libarchive}
  with the \emph{alternative} order setting.
  All data unit is \%. (total 100\%). }
  \label{f:autofuzz-libarchive-alt}
\end{figure}

\PP{Case study: adding good fuzzer (5-fuzzers).}
In the following case studies, we further explain
how \sys can exfiltrate well-performing fuzzers
from various fuzzer sets.
When it comes to 4-fuzzer and 5-fuzzer cases
in the \emph{alternative} order setting in \autoref{f:fuzzer-num-libarchive},
we see huge coverage increases in \sys and \sysm
as a result of introducing one more good fuzzer,
\libfuzzer (ranked at \nth{3}).
However, \sys experiences a much higher performance increase
compared to \sysm.
We provide a reasonable explanation with~\autoref{f:autofuzz-libarchive-alt5}
focusing on the resource allocation details of \sys.
First,
\sys rarely allocates resources
to poorly performing fuzzers (i.e., \radamsa and \aflfast).
Note that \sys consumed only 6\% and 2\% of resources
for the bad fuzzers during the entire focus phases.
Even after taking the preparation phases into account,
\sys spent only 22\% resources total
for the two bad fuzzers.
However, \sysm equally distributed the resources, 20\% for each fuzzer, regardless of their performance.
Therefore, \sysm spent 18\% more resources
on the poorly performing fuzzers.
Second,
\sys allocated more resources toward
three well-performing fuzzers (i.e., \redqueen, \lafintel,
\libfuzzer), which boosts the performance.
The first pie chart shows that
\sys respectively allocated
35\%, 30\%, and 27\% of resources to them
during the entire focus phases.
Note that more resources were allotted to
better-performing fuzzers
even though \sys does not have any prior knowledge about
which fuzzers are in what ranks.
However,
\sysm equally assigns 20\% of resources to
each of the three fuzzers.
For total resources
spent for those three fuzzers,
\sys and \sysm consumed 78\% and 60\%, respectively.

\PP{Case study: adding bad fuzzer (6-fuzzers)}
Contrary to the previous case study,
we explain how \sys and \sysm
react when a poorly performing fuzzer is introduced
to their baseline.
As described in~\autoref{f:autofuzz-libarchive-alt},
we added one more fuzzer \afl
(ranked at \nth{9}) and captured
how the \ignore{bitmap} coverage changes.
After the addition,
both \sys and \sysm had a performance drop.
The resource allocation detail is shown in \autoref{f:autofuzz-libarchive-alt6}.
\sys allocates only 2\% of resources to \afl
during the entire focus phase,
which demonstrates that it can exfiltrate
poorly performing fuzzers properly.
However, when it comes to overall resources
spent for \afl, it increases to 8\%,
which is the major reason for the slowdown.
In this example,
\sys spent 6\% of resources
to capture possible trend changes
during the preparation phases.
Note that this cost is unavoidable.
Therefore,
as more fuzzers are introduced,
the resources spent for the preparation phases
can become a burden for \sys.
However, the captured runtime trend
more than compensates for the slowdown incurred by
the preparation phases.
As a consequence,
\sys still outperforms \sysm
after introducing \afl,
which demonstrates that
the resource allocation algorithm of \sys
can highlight good fuzzers.

\subsection{Bug Discovery Evaluation}
\label{ss:eval-bug-discovery}
To compare the number of bugs found by \sys and individual fuzzers,
we run ASan-instrumented \fts and \unifuzz targets and collect generated crashes.
We triage the crashes and measure deduplicated bugs
by taking the top three stack frames and
using them as unique identifiers for the bugs.
\sys found the most bugs on average compared to individual fuzzers, \cupid, and \enfuzz.
Moreover, when we aggregate the results of all 10 fuzzing repetitions,
\sys still uncovered the most bugs,
some of which are not found by any individual fuzzer.
The detailed results are in \autoref{ss:bug-count}.

\subsection{Evaluation of \sysn decision}
\label{ss:eval-decision}
In this section,
we evaluate
how accurate the decisions of \sys are
in terms of resource allocation \ignore{at runtime}.
Resource allocation is the essence of \sys
because it reflects the trends of baseline fuzzers
and makes a subset of baseline fuzzers to be online
predicted to be efficient for the current workload.
If prediction about runtime trends
is inaccurate and does not continue,
less coverage will be explored.
Therefore, we can estimate
the continuity of measured trends as
the end result achieved
as a result of deploying specific resource
allocation per round.
However, it is not practical to compare \sys
with all possible resource allocation decisions
because it is infinite.
To this end, we carefully design the following evaluation.

\begin{enumerate}[label*=\arabic*.]
\item For each round,
    record when the preparation phase ended ($time_{prep\_end}$) and
    total time budget assigned for the focus phase ($time_{focus\_total}$).
    Note that the value of these two variables
    can be different in every round
    based on when the early exit occurs.
\item Collect all output corpus created
    earlier than $time_{prep\_end}$.
    We call it \emph{snapshot}.
    The snapshot restores the full status
    of the fuzzing campaign
    to the state right before the focus phase.
\item For the number of baseline fuzzers, repeat the below procedures.
  \begin{enumerate}[label*=\arabic*.]
    \item Select one fuzzer and
    update resource allocation so that
    all resources are assigned to that fuzzer.
    \item Run the fuzzer for $time_{focus\_total}$
    and measure the coverage.
    Note that each individual fuzzer run represents
    synthetic decisions made for comparison.
    \item Restore the current state to right before the focus phase using the \emph{snapshot}.
  \end{enumerate}

\item Lastly, run fuzzers
    selected by the resource allocation of \sys
    for $time_{focus\_total}$ and measure the coverage.
\end{enumerate}

\begin{figure*}[t]
\begin{subfigure}[t]{0.5\textwidth}
    \centering
    \includegraphics[width=\columnwidth]
    {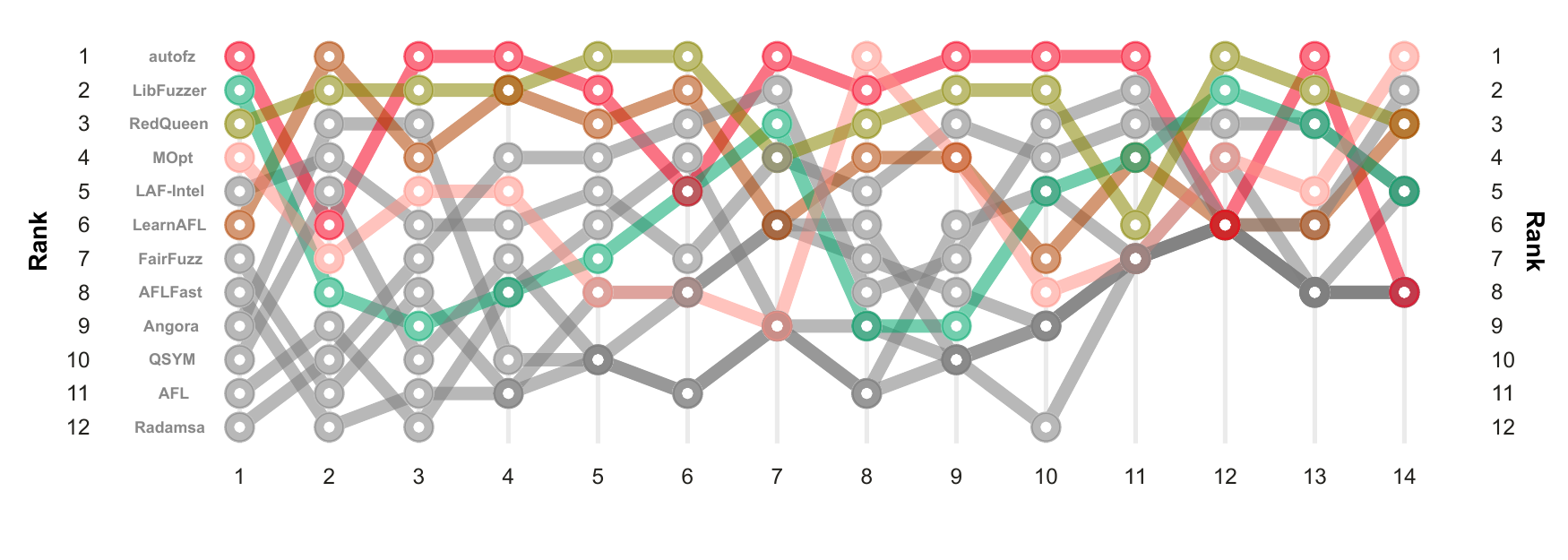}\\
    \vspace{-0.3cm}
    \includegraphics[width=\columnwidth]
    {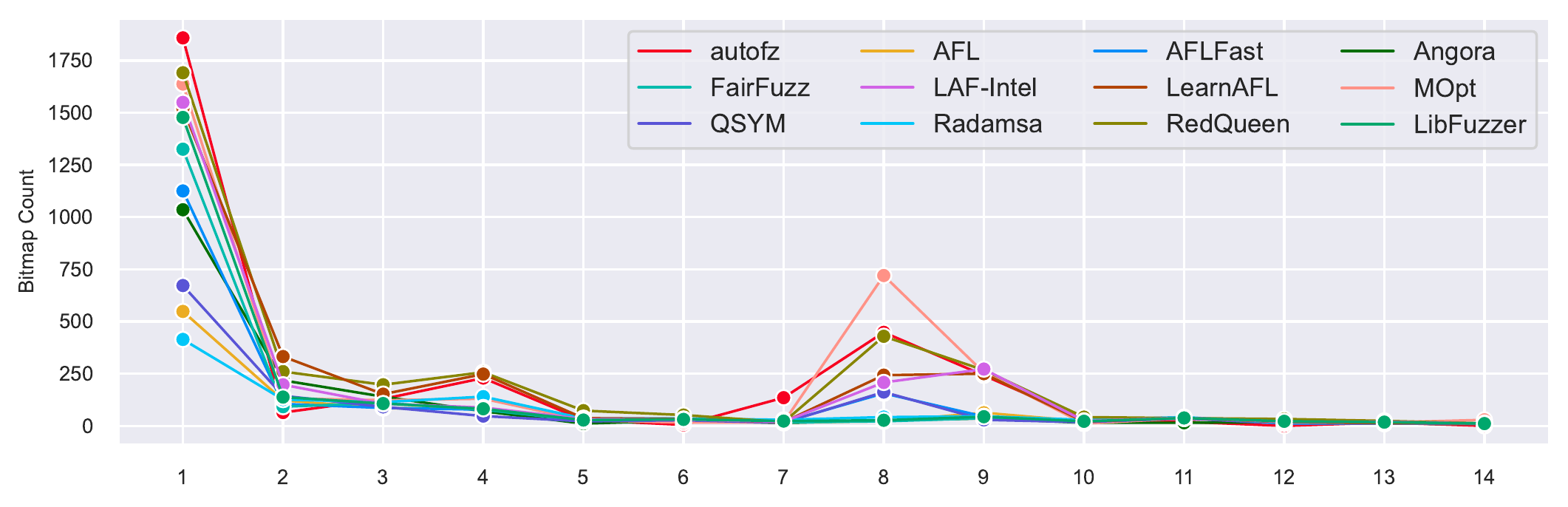}
    \includegraphics[width=\columnwidth]{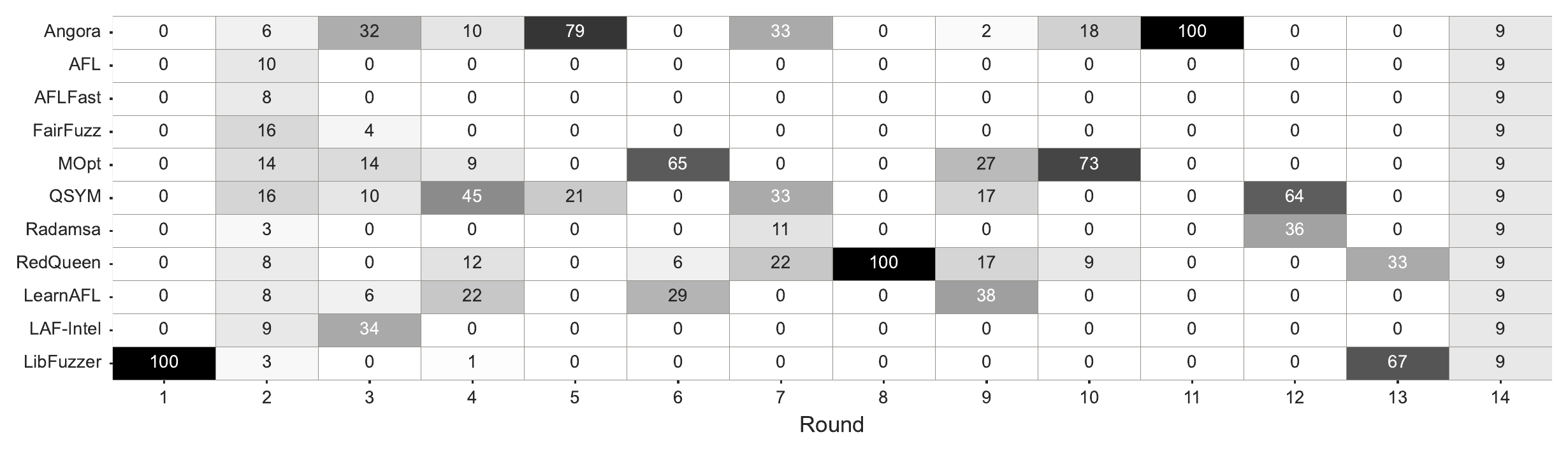}
    \caption{Evaluation of \sys decision on \libarchive}
  \label{f:eval-round-libarchive}
\end{subfigure}
\begin{subfigure}[t]{0.5\textwidth}
    \centering
    \includegraphics[width=\columnwidth]
    {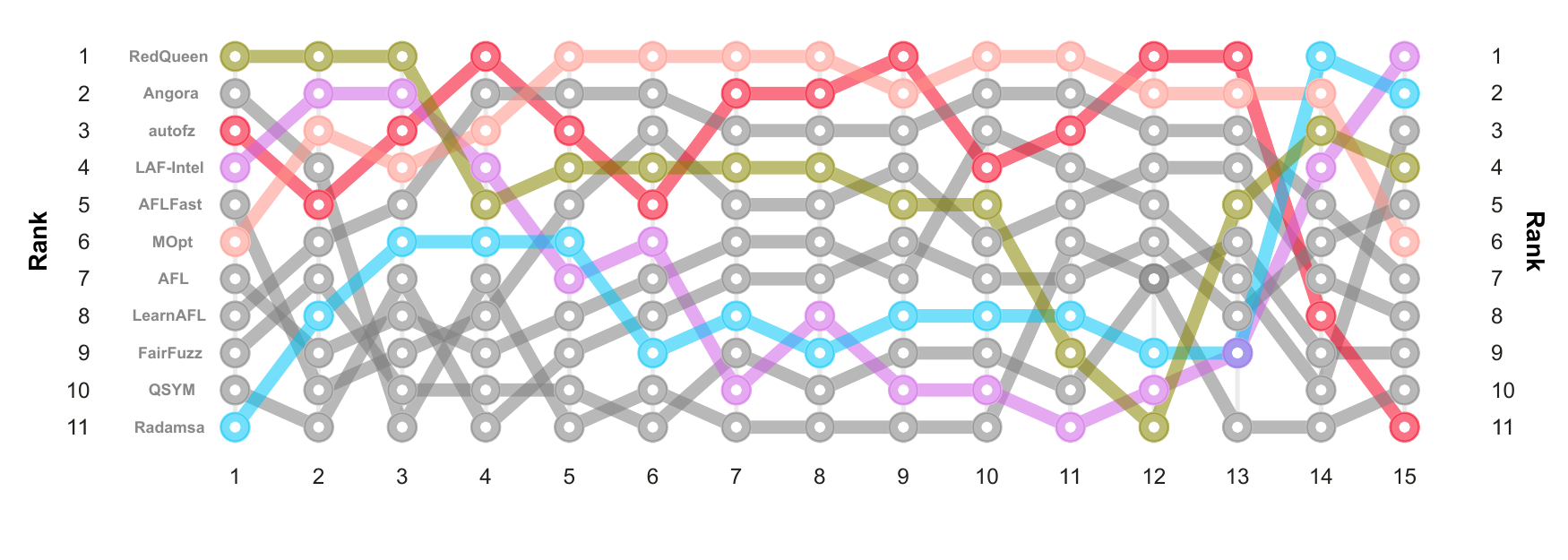}\\
    \vspace{-0.3cm}
    \includegraphics[width=\columnwidth]
    {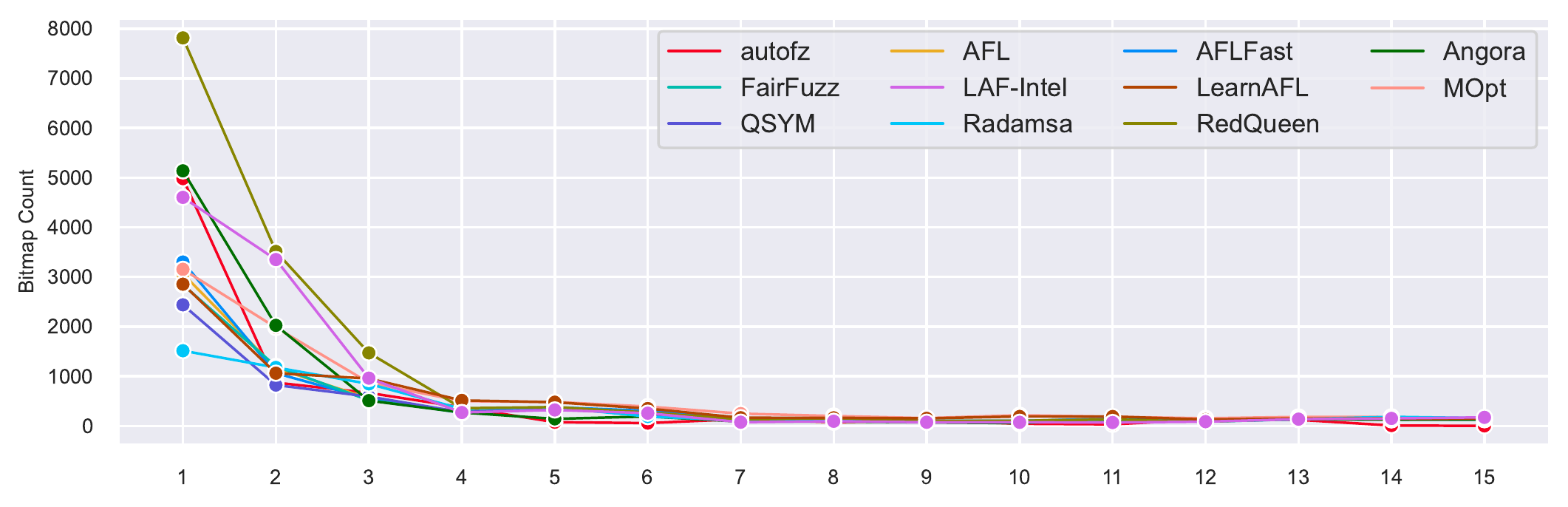}
    \includegraphics[width=\columnwidth]
    {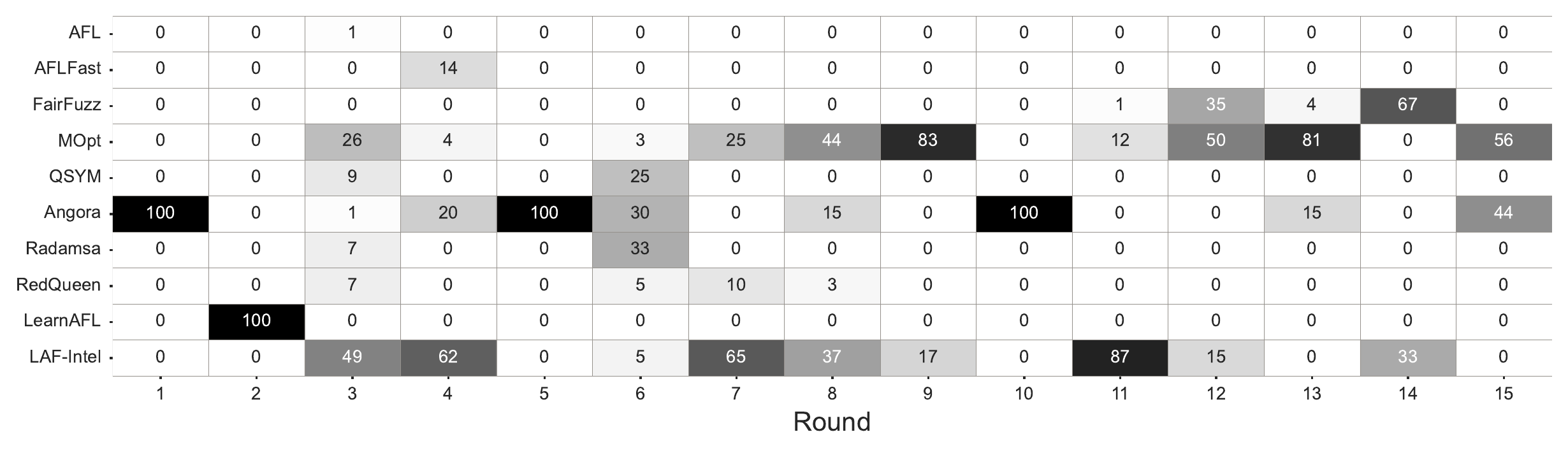}
    \caption{Evaluation of \sys decision on exiv2}
  \label{f:eval-round-exiv2}
\end{subfigure}
\caption{Comparison between resource allocation decision of \sys (\autoref{a:resource-assignment}) and decisions
allocating entire resources to each baseline fuzzers.
The bump chart compares the ranks of all decisions
in terms of \afl bitmap coverage.
We highlight decisions that have ranked top at least once.
The line plots illustrate bitmap counts
of different decisions at each round.
The heatmap presents
\sys's detailed resource allocation decisions.
Note that the allocation reflects
runtime trends of different fuzzer
at each round.
}
\end{figure*}

The evaluation results on
\libarchive and \exiv2
are presented in
\autoref{f:eval-round-libarchive} and
\autoref{f:eval-round-exiv2}, respectively.
As shown in \autoref{f:eval-round-libarchive},
the decision made by \sys allows it to take first place
eight times over 14 rounds and rank 2.64 on average.
The \exiv2 target,
presented in \autoref{f:eval-round-exiv2},
ranks 3.5 on average and
took first place four times over 15 rounds.
This result demonstrates that
the decision of \sys is effective
compared with others and proves that
the captured trends maintain
until the next measurement.

\PN{Saturation and effectiveness of \sys}
The result shows that the allocation decisions of \sys
become less efficient
as the number of new coverage reported
by the baseline fuzzer saturates.
Although \sys ranks higher on average
in fuzzing \libarchive compared to \exiv2,
note that \afl bitmap count saturates faster
in \exiv2 than in \libarchive.
In detail, after round 4 in \exiv2,
all different decisions including \sys
do not exhibit meaningful bitmap coverage increases.
Therefore, no matter which fuzzers are activated as a result of different resource allocations,
it is difficult to make progress after round 4.
For example,
\sys ranked 11 in round 15,
but the bitmap density difference
between \lafintel (the best) and \sys (the worst)
is only 0.07\%.

\PP{Noise introduced by inherent randomness.}
Although we carefully designed our evaluation
so that all different decisions have
identical starting lines using \emph{snapshot} every round,
it is impossible to completely prevent introducing noise
due to the inherent randomness in fuzzing campaigns.
For example, in \autoref{f:eval-round-libarchive},
\sys allocates all resources to \libfuzzer
at the first round,
making \sys's decision identical to the \libfuzzer case.
However, the result of following \sys's decision (ranked \nth{1})
slightly outperforms the \libfuzzer decision (ranked \nth{2}),
as shown in \autoref{f:eval-round-libarchive}.
We believe that the inherent randomness,
such as in input mutations,
causes this difference, which we call noise.

\PP{Seed synchronization might change the trends.}
We can explain
why the captured trends sometimes will not be sustained
in two aspects.
First, the preparation phase was not
long enough to capture the trends properly.
Second, \sys captured the trends accurately,
but the seed synchronization after the preparation
allows the other individual fuzzers to report
strong trends.
For example,
we can observe that
\redqueen does not perform well
compared to \angora
when each fuzzer runs individually,
highlighted in \autoref{eval:resource distribution}.
The first round of the preparation on \exiv2
reports that \angora exhibits the strongest trend,
which follows the observations.
Therefore,
\sys allocates all resources to \angora
(see the heatmap presented in \autoref{f:eval-round-exiv2}).
However, our evaluation revealed that
\redqueen performed the best
during the first round of the focus phase (see the second graph of \autoref{f:eval-round-exiv2}).
Note that this result does not follow our initial observation.
We believe that this happens because
the first round of the preparation phase explored
paths that used to be challenging
for \redqueen to resolve.
Therefore, after the seed synchronization,
\redqueen does not need to spend its resources
to reach those paths anymore and
exhibit the strongest trends
during the focus phase.
However, note that \angora still performs well
in the first focus phase
even though it cannot be ranked on top.
This demonstrates that
runtime trends captured in the preparation phase
are strong enough to allow \sys to achieve good performance.

\section{Discussion}
\label{s:discussion}
We discuss the limitations and future works of \sys.

\PN{Metric diversity.}
\sys utilizes \afl bitmap to compare runtime trends,
which favors the fuzzers that seek to maximize path coverage.
While the coverage is the most popular and explicit indicator of progress in fuzzing,
relying on a single metric can potentially lead to unfair comparison
with various fuzzers utilizing metrics other than the coverage
\cite{wang19:analyze-coverage, fioraldi20:likely, wang20:not, wen20:memlock, petsios17:slowfuzz}.
Therefore, supporting multiple metrics
besides path coverage can achieve fairness and better
efficiency in terms of resource allocation.
For example, different metrics can be used to break the tie,
especially when one metric is saturated.

\PN{Bad fuzzer elimination.}
A particular fuzzer might not perform well
throughout the fuzzing \ignore{campaign}
for specific targets
(e.g., \radamsa on \exiv2 shown in \autoref{f:autofuzz-eval-best}).
\sys automatically prevents poorly performing fuzzers
from being online during the focus phase
through resource allocation.
However, the same amount of resources
are allocated to all baseline fuzzers
during the preparation phase
to measure runtime trends.
If poorly performing fuzzers can be eliminated
from the baseline timely,
\sys can achieve better resource utilization.

\section{Conclusion}
\label{s:conclusion}
This paper presented \sys,
a meta-fuzzer providing 
fine-grained and non-intrusive fuzzer 
orchestration.
Our evaluation result illustrates that 
automated fuzzer composition
without any prior knowledge is effective.
By observing the trends of fuzzers at runtime 
and distributing the computing resources properly, 
\sys has not only beat the individual fuzzers 
but also the state-of-the-art collaborative fuzzing approaches.
We expect that  
\sys can \emph{bridge the gap 
between developing new fuzzers 
and their effective deployment.}

\section*{Acknowledgment}
\label{s:ack}
We thank the anonymous reviewers for their insightful feedback.
This research was supported, in part, by Cisco,
the NSF award CNS-1749711,
ONR under grant N00014-23-1-2095,
DARPA V-SPELL N66001-21-C-402,
and gifts from Facebook, Mozilla, Intel, VMware and Google.

\bibliographystyle{IEEEtranS}
\bibliography{p,sslab,fuzz,conf}

\begin{thebibliography}{10}
\providecommand{\url}[1]{#1}
\csname url@samestyle\endcsname
\providecommand{\newblock}{\relax}
\providecommand{\bibinfo}[2]{#2}
\providecommand{\BIBentrySTDinterwordspacing}{\spaceskip=0pt\relax}
\providecommand{\BIBentryALTinterwordstretchfactor}{4}
\providecommand{\BIBentryALTinterwordspacing}{\spaceskip=\fontdimen2\font plus
\BIBentryALTinterwordstretchfactor\fontdimen3\font minus
  \fontdimen4\font\relax}
\providecommand{\BIBforeignlanguage}[2]{{%
\expandafter\ifx\csname l@#1\endcsname\relax
\typeout{** WARNING: IEEEtranS.bst: No hyphenation pattern has been}%
\typeout{** loaded for the language `#1'. Using the pattern for}%
\typeout{** the default language instead.}%
\else
\language=\csname l@#1\endcsname
\fi
#2}}
\providecommand{\BIBdecl}{\relax}
\BIBdecl

\bibitem{lafintel}
``Circumventing fuzzing roadblocks with compiler transformations,'' 2016,
  \url{https://lafintel.wordpress.com/}.

\bibitem{autofz-extended}
\BIBentryALTinterwordspacing
Anonymous, ``\sys: Automated fuzzer composition at runtime,'' 2023. [Online].
  Available: \url{https://arxiv.org/abs/xxxxxxxx}
\BIBentrySTDinterwordspacing

\bibitem{arcuri12:random-testing}
A.~Arcuri, M.~Z. Iqbal, and L.~Briand, ``Random testing: Theoretical results
  and practical implications,'' \emph{IEEE Transactions on Software
  Engineering}, 2012.

\bibitem{aschermann19:redqueen}
C.~Aschermann, S.~Schumilo, T.~Blazytko, R.~Gawlik, and T.~Holz, ``{REDQUEEN:}
  fuzzing with input-to-state correspondence,'' in \emph{NDSS}, 2019.

\bibitem{boehme21:fuzzing-challenge}
M.~Boehme, C.~Cadar, and A.~ROYCHOUDHURY, ``Fuzzing: Challenges and
  reflections,'' \emph{IEEE Software}, vol.~38, 2021.

\bibitem{bohme20:entropy}
M.~B{\"{o}}hme, V.~J.~M. Man{\`{e}}s, and S.~K. Cha, ``Boosting fuzzer
  efficiency: an information theoretic perspective,'' in \emph{ESEC/FSE}, 2020.

\bibitem{cgf-bohme-ccs16}
M.~B{\"o}hme, V.-T. Pham, and A.~Roychoudhury, ``{Coverage-based Greybox
  Fuzzing as Markov Chain},'' in \emph{CCS}, 2016.

\bibitem{bohme19:cover-based-greyb-fuzzin-markov-chain}
M.~B{\"{o}}hme, V.~Pham, and A.~Roychoudhury, ``Coverage-based greybox fuzzing
  as markov chain,'' \emph{{IEEE} Trans. Software Eng.}, vol.~45, 2019.

\bibitem{cadar:klee08}
C.~Cadar, D.~Dunbar, and D.~Engler, ``{KLEE: Unassisted and Automatic
  Generation of High-coverage Tests for Complex Systems Programs},'' in
  \emph{OSDI}, 2008.

\bibitem{chen2018angora}
P.~Chen and H.~Chen, ``Angora: Efficient fuzzing by principled search,'' in
  \emph{S\&P Oakland}, 2018.

\bibitem{chen19:enfuz}
Y.~Chen, Y.~Jiang, F.~Ma, J.~Liang, M.~Wang, C.~Zhou, X.~Jiao, and Z.~Su,
  ``Enfuzz: Ensemble fuzzing with seed synchronization among diverse fuzzers,''
  in \emph{Security}, 2019.

\bibitem{aimd}
D.-M. Chiu and R.~Jain, ``Analysis of the increase and decrease algorithms for
  congestion avoidance in computer networks,'' \emph{Computer Networks and ISDN
  systems}, vol.~17, 1989.

\bibitem{darpa:cgc}
DARPA, ``{DARPA Cyber Grand Challenge Final Event Archive},'' 2016,
  \url{https://www.lungetech.com/cgc-corpus/}.

\bibitem{demvsar:cd}
J.~Dem{\v{s}}ar, ``Statistical comparisons of classifiers over multiple data
  sets,'' \emph{The Journal of Machine Learning Research}, vol.~7, 2006.

\bibitem{ding21:empir-study-oss-fuzz-bugs}
Z.~Ding and C.~L. Goues, ``An empirical study of oss-fuzz bugs,'' in
  \emph{IEEE/ACM 18th International Conference on Mining Software Repositories
  (MSR)}, 2021.

\bibitem{dolan-gavitt16:lava}
B.~Dolan{-}Gavitt, P.~Hulin, E.~Kirda, T.~Leek, A.~Mambretti, W.~K. Robertson,
  F.~Ulrich, and R.~Whelan, ``{LAVA:} large-scale automated vulnerability
  addition,'' in \emph{S\&P Oakland}, 2016.

\bibitem{fioraldi20:likely}
A.~Fioraldi, D.~C. D'Elia, and D.~Balzarotti, ``The use of likely invariants as
  feedback for fuzzers,'' in \emph{Security}, 2021.

\bibitem{fioraldi20:afl++}
A.~Fioraldi, D.~Maier, H.~Ei{\ss}feldt, and M.~Heuse, ``{AFL++} : Combining
  incremental steps of fuzzing research,'' in \emph{{USENIX} Workshop on
  Offensive Technologies ({WOOT})}, 2020.

\bibitem{clusterfuzz-google-web}
Google, ``{Fuzzing for Security},'' 2012,
  \url{https://blog.chromium.org/2012/04/fuzzing-for-security.html}.

\bibitem{oss-fuzz-google-web}
------, ``{OSS-Fuzz - Continuous Fuzzing for Open Source Software},'' 2016,
  \url{https://github.com/google/oss-fuzz}.

\bibitem{honggfuzz-found-cve}
------, ``{Honggfuzz Found Bugs},'' 2018,
  \url{https://github.com/google/honggfuzz\#trophies}.

\bibitem{fts}
------, ``{Fuzzer Test Suite},'' 2021,
  \url{https://opensource.google/projects/fuzzer-test-suite}.

\bibitem{guler20:cupid}
E.~G{\"{u}}ler, P.~G{\"{o}}rz, E.~Geretto, A.~Jemmett, S.~{\"{O}}sterlund,
  H.~Bos, C.~Giuffrida, and T.~Holz, ``Cupid : Automatic fuzzer selection for
  collaborative fuzzing,'' in \emph{ACSAC}, 2020.

\bibitem{hazimeh21:magma}
A.~Hazimeh, A.~Herrera, and M.~Payer, ``Magma: {A} ground-truth fuzzing
  benchmark,'' in \emph{{SIGMETRICS}}, 2021.

\bibitem{radamsa}
A.~Helin, ``{Radamsa},'' 2021, \url{https://gitlab.com/akihe/radamsa}.

\bibitem{langfuzz-holler-sec12}
C.~Holler, K.~Herzig, and A.~Zeller, ``{Fuzzing with Code Fragments.}'' in
  \emph{Security}, 2012.

\bibitem{cab-fuzz-kim-atc17}
S.~Y. Kim, S.~Lee, I.~Yun, W.~Xu, B.~Lee, Y.~Yun, and T.~Kim, ``{CAB-Fuzz:
  Practical Concolic Testing Techniques for COTS Operating Systems},'' in
  \emph{ATC}, 2017.

\bibitem{klees18:evaluat-fuzz-testin}
G.~Klees, A.~Ruef, B.~Cooper, S.~Wei, and M.~Hicks, ``Evaluating fuzz
  testing,'' in \emph{CCS}, 2018.

\bibitem{lemieux18:fairfuzz}
C.~Lemieux and K.~Sen, ``Fairfuzz: a targeted mutation strategy for increasing
  greybox fuzz testing coverage,'' in \emph{ASE}, 2018.

\bibitem{li21:unifuzz}
Y.~Li, S.~Ji, Y.~Chen, S.~Liang, W.-H. Lee, Y.~Chen, C.~Lyu, C.~Wu, R.~Beyah,
  P.~Cheng, K.~Lu, and T.~Wang, ``{UNIFUZZ}: A holistic and pragmatic
  metrics-driven platform for evaluating fuzzers,'' in \emph{Security}, 2021.

\bibitem{liang18:pafl}
J.~Liang, Y.~Jiang, Y.~Chen, M.~Wang, C.~Zhou, and J.~Sun, ``Pafl: Extend
  fuzzing optimizations of single mode to industrial parallel mode,'' in
  \emph{ESEC/FSE}, 2018.

\bibitem{lyu19:mopt}
C.~Lyu, S.~Ji, C.~Zhang, Y.~Li, W.~Lee, Y.~Song, and R.~Beyah, ``{MOPT:}
  optimized mutation scheduling for fuzzers,'' in \emph{Security}, 2019.

\bibitem{manes:tse:2021}
V.~J.~M. Man{\`{e}}s, H.~Han, C.~Han, S.~K. Cha, M.~Egele, E.~J. Schwartz, and
  M.~Woo, ``The art, science, and engineering of fuzzing: A survey,''
  \emph{IEEE Transactions on Software Engineering}, vol.~47, 2021.

\bibitem{fuzzbench}
J.~Metzman, L.~Szekeres, L.~M.~R. Simon, R.~T. Sprabery, and A.~Arya,
  ``Fuzzbench: An open fuzzer benchmarking platform and service,'' in
  \emph{ESEC/FSE}, 2021.

\bibitem{microsoft-onefuzz}
Microsoft, ``{Fuzzing for Security},'' 2020,
  \url{https://www.microsoft.com/en-us/research/project/project-onefuzz/}.

\bibitem{miller1990}
B.~P. Miller, L.~Fredriksen, and B.~So, ``{An Empirical Study of the
  Reliability of UNIX Utilities},'' \emph{Commun. ACM}, vol.~33, 1990.

\bibitem{profuzzbench}
R.~Natella and V.-T. Pham, ``Profuzzbench: A benchmark for stateful protocol
  fuzzing,'' in \emph{ISSTA}, 2021.

\bibitem{cgroups}
C.~L. Paul~Menage, Paul~Jackson, ``Control groups,'' 2022,
  \url{https://www.kernel.org/doc/html/latest/admin-guide/cgroup-v1/cgroups.html}.

\bibitem{petsios17:slowfuzz}
T.~Petsios, J.~Zhao, A.~D. Keromytis, and S.~Jana, ``Slowfuzz: Automated
  domain-independent detection of algorithmic complexity vulnerabilities,'' in
  \emph{CCS}, 2017.

\bibitem{afl-found-cve}
M.~Rash, ``{A Collection of Vulnerabilities Discovered by the AFL Fuzzer},''
  2017, \url{https://github.com/mrash/afl-cve}.

\bibitem{vuzzer-rawat-ndss17}
S.~Rawat, V.~Jain, A.~Kumar, L.~Cojocar, C.~Giuffrida, and H.~Bos, ``{VUzzer:
  Application-aware Evolutionary Fuzzing},'' in \emph{NDSS}, 2017.

\bibitem{serebryany15:libfuzzer}
K.~Serebryany, ``libfuzzer a library for coverage-guided fuzz testing,''
  \emph{LLVM project}, 2015.

\bibitem{driller-stephens-ndss16}
N.~Stephens, J.~Grosen, C.~Salls, A.~Dutcher, R.~Wang, J.~Corbetta,
  Y.~Shoshitaishvili, C.~Kruegel, and G.~Vigna, ``{Driller: Augmenting Fuzzing
  through Selective Symbolic Execution},'' in \emph{NDSS}, 2016.

\bibitem{swiecki21:honggfuzz}
R.~Swiecki, ``Honggfuzz,'' 2021, \url{http://code.google.com/p/honggfuzz}.

\bibitem{syzkaller-found-cve}
Syzkaller, ``{Syzkaller Found Bugs - Linux Kernel},'' 2018,
  \url{https://github.com/google/syzkaller/blob/master/docs/linux/found_bugs.md}.

\bibitem{wang19:analyze-coverage}
J.~Wang, Y.~Duan, W.~Song, H.~Yin, and C.~Song, ``Be sensitive and
  collaborative: Analyzing impact of coverage metrics in greybox fuzzing,'' in
  \emph{RAID}, 2019.

\bibitem{wang20:not}
Y.~Wang, X.~Jia, Y.~Liu, K.~Zeng, T.~Bao, D.~Wu, and P.~Su, ``Not all coverage
  measurements are equal: Fuzzing by coverage accounting for input
  prioritization.'' in \emph{NDSS}, 2020.

\bibitem{wen20:memlock}
C.~Wen, H.~Wang, Y.~Li, S.~Qin, Y.~Liu, Z.~Xu, H.~Chen, X.~Xie, G.~Pu, and
  T.~Liu, ``Memlock: Memory usage guided fuzzing,'' in \emph{ICSE}, 2020.

\bibitem{xu:os-fuzz}
W.~Xu, S.~Kashyap, C.~Min, and T.~Kim, ``{Designing New Operating Primitives to
  Improve Fuzzing Performance},'' in \emph{CCS}, 2017.

\bibitem{learnafl}
T.~Yue, Y.~Tang, B.~Yu, P.~Wang, and E.~Wang, ``Learnafl: Greybox fuzzing with
  knowledge enhancement,'' \emph{IEEE Access}, vol.~7, 2019.

\bibitem{yun18:qsym}
I.~Yun, S.~Lee, M.~Xu, Y.~Jang, and T.~Kim, ``{QSYM} : {A} practical concolic
  execution engine tailored for hybrid fuzzing,'' in \emph{Security}, 2018.

\bibitem{afl-zalewski-web}
M.~Zalewski, ``{American Fuzzy Lop (2.52b)},'' 2018,
  \url{http://lcamtuf.coredump.cx/afl/}.

\bibitem{afl-detail}
------, ``{AFL} technique details,'' 2022,
  \url{https://github.com/google/AFL/blob/master/docs/technical_details.txt}.

\bibitem{osterlund21:collab}
S.~Österlund, E.~Geretto, A.~Jemmett, E.~Güler, P.~Görz, T.~Holz,
  C.~Giuffrida, and H.~Bos, ``Collabfuzz: A framework for collaborative
  fuzzing,'' in \emph{Proceedings of the 14th European Workshop on Systems
  Security}, 2021.

\end{thebibliography}

\appendix
\section{Appendix}
\begin{minipage}{\columnwidth}
\begin{algorithm}[H]
  \caption{Two Phase Algorithm}
  \label{a:two-phase-combined}
  \small
  \begin{algorithmic}[1]
    \Input
          \State $\mathbb{F} \gets \{{f_{1},f_{2},...,f_{n}\}}$, $f_{n}$ is instance of $n_{th}$ baseline fuzzer
          \State $\mathbb{B} \gets \{{b_{1},b_{2},...,b_{n}\}}$, $b_{n}$ is bitmap of $f_{n}$
          \State $T_{prep,focus} \gets$ Time budget for preparation/focus phase
          \State $\theta_{init,cur} \gets$ Initial and current threshold
        \State $C \gets$ How many CPU cores are assigned
    \EndInput
          \Function{autofz_main}{$\mathbb{F}$, $T_{prep}$, $T_{focus}$, $\theta_{cur}=\theta_{init}$, $C$}
    \While {NOT timeout}
          \State \Call{seed\_sync}{$\mathbb{F}$}
          \State {{$Exit_{early}$, $T_{remain}$ $\gets$ \Call{prep\_phase}{$\mathbb{F}$, $\mathbb{B}$, $T_{prep}$, $\theta_{cur}$, $C$}}}
          \State $\mathbb{RA} \gets \Call{resource\_allocator}{\mathbb{F}, \mathbb{B}, Exit_{early}}$
      \State $\theta_{cur} \gets$ $(Exit_{early})$ $?$ $\theta_{cur}+\theta_{init}$ $:$ $\theta_{cur}*0.5$
      \Comment{AIMD}
      \label{lst:two-phase-aimd}
          \State \Call{seed\_sync}{$\mathbb{F}$}
          \State \Call{focus\_phase}{$\mathbb{F}$, $\mathbb{RA}$, $T_{focus} + T_{remain}$, $C$}
    \EndWhile
    \EndFunction
  \end{algorithmic}
\end{algorithm}
\begin{algorithm}[H]
  \small
  \caption{Focus Phase}
  \label{a:focus-phase}
  \begin{algorithmic}[1]
    \Function{focus\_phase}{$\mathbb{F}$, $\mathbb{RA}$, $T_{focus}$, $C$}
   \If {$C == 1$}
        \State $T_{focus\_total} \gets T_{focus} \times \Call{num\_of\_fuzzers}{\mathbb{F}}$ \label{lst:focus-total}
        \State {{ $\mathbb{F}$ $\gets$ \Call{sort\_fuzzers}{$\mathbb{F}$, key = $\mathbb{RA}$, order = descending}}} \label{lst:sort-fuzzers}
        \For {\textbf{each} $f$ $\in$ $\mathbb{F}$}
    	  \State $T_{run} \gets T_{focus\_total} \times \mathbb{RA}[f]$ \label{lst:time-budget}
          \If{$T_{run} > 0$} \label{lst:focus-run1}
            \State $\Call{run\_fuzzer}{f, T_{run}}$ \label{lst:focus-run2}
            \State $\Call{seed\_sync}{\mathbb{F}}$ \label{lst:focus-seed-sync}
          \EndIf
        \EndFor
    \Else \Comment{multi-core implementation}
      \State $c \gets C \times \mathbb{RA}[f]$
      \label{lst:focus-multi-core-cal}
      \State \Call{run\_fuzzers\_parallel\_seed\_sync}{$\mathbb{F}$, $T_{focus}$, $c$}
      \label{lst:focus-multi-core}
      \State $\Call{seed\_sync}{\mathbb{F}}$
    \EndIf
    \EndFunction
  \end{algorithmic}
\end{algorithm}
\end{minipage}

\subsection{Queue Size Statistics And Observation}
\begin{figure*}
\begin{minipage}{\textwidth}
\begin{figure}[H]
  \begin{center}
    {
    \fontsize{6pt}{6pt}\selectfont
     \def\svgwidth{\textwidth}
     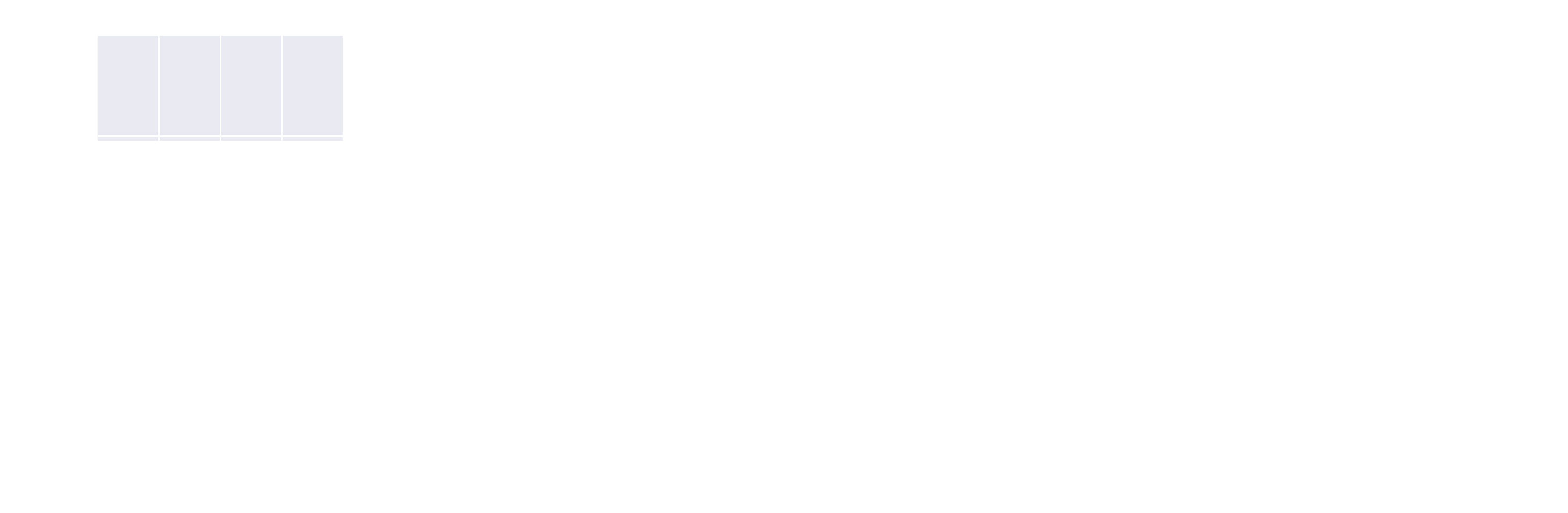
    }
  \end{center}
  \vspace{-0.45cm}
	\caption{Evaluation of \sys on \unifuzz and \fts in terms of \textbf{line coverage}.
 }
  \label{f:autofuzz-eval-line}
\end{figure}

\begin{figure}[H]
  \begin{center}
    {
    \fontsize{6pt}{6pt}\selectfont
     \def\svgwidth{\textwidth}
     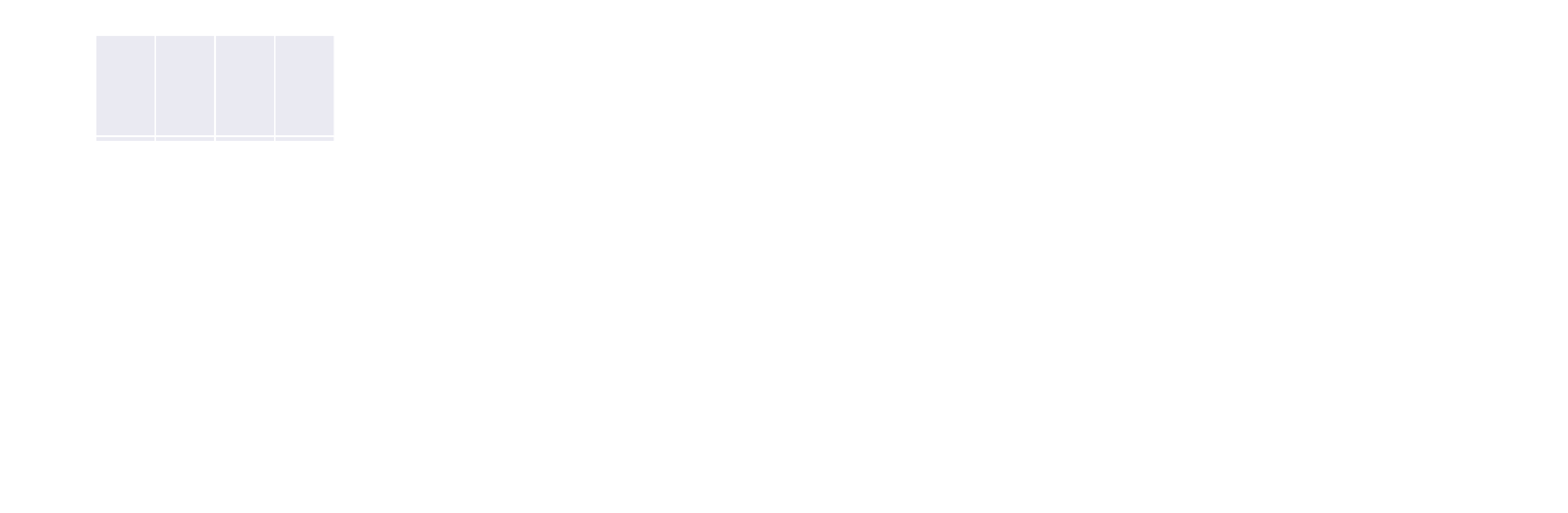
    }
  \end{center}
  \vspace{-0.45cm}
	\caption{Evaluation of \sys on \unifuzz and \fts in terms of \textbf{queue size}.
 }
  \label{f:autofuzz-eval-queue}
\end{figure}
\end{minipage}
\end{figure*}

\PP {Queue size measurement.}
We collect \ignore{all} files in \cc{queue} output directories
to measure changes in queue size over 24 hours.
Whenever an \emph{interesting seed} is found,
each fuzzer stores the seed as a file
in the queue directory.
Each fuzzer has a different guideline to determine
which seed is interesting, for example,
based on whether a seed explores new coverage.
\sys has no clear definition of interesting seeds,
so it collects all interesting seeds
from the queue directories of all baseline fuzzers
and deduplicates the same seeds by comparing their hash.
\autoref{f:autofuzz-eval-queue} presents the queue size change
over 24 hours for \sys and individual fuzzers on \unifuzz and \fts.

\PP{Observation}
We found that \angora significantly outperforms other fuzzers, including \sys,
in terms of queue size on \freetype2, \pdftotext, \mujs, and \tcpdump.
However, \angora embarrassingly underperforms
in terms of line and edge coverage
on those targets except \tcpdump.
We analyzed the logs of \angora and found that,
especially when the coverage saturates,
it generated lots of new seeds
contributing a tiny fraction of the \emph{context-sensitive} coverage bitmap.
For example, in \mujs,
\angora generated around \num{80000} interesting seeds
and achieved 99.98\% bitmap coverage.
However, it requires around \num{90000} interesting seeds
to cover the remaining 0.02\% bitmap coverage.

\PP{Possible interpretation}
We speculate that abundant seeds hinder \angora from
mutating the seeds that could be meaningful
in terms of other metrics.
For example, such high bitmap density can lead to
a higher rate of hash collision and prevent
interesting seeds in terms of different metrics
from being generated.
As a result, \angora underperforms on \freetype2, \pdftotext, \mujs
in terms of AFL bitmap density, edge coverage, and bug-finding
because the context-sensitive coverage bitmap of \angora saturates early.
On the other hand, for \tcpdump,
\angora outperforms others
not only in queue size but also in bug finding.
This result implies that
considering multiple metrics in \sys
could be beneficial in terms of finding bugs.

\clearpage

\begin{minipage}{\textwidth}
\begin{minipage}{0.45\textwidth}
\begin{figure}[H]
  \begin{center}
    \includegraphics[width=\columnwidth]{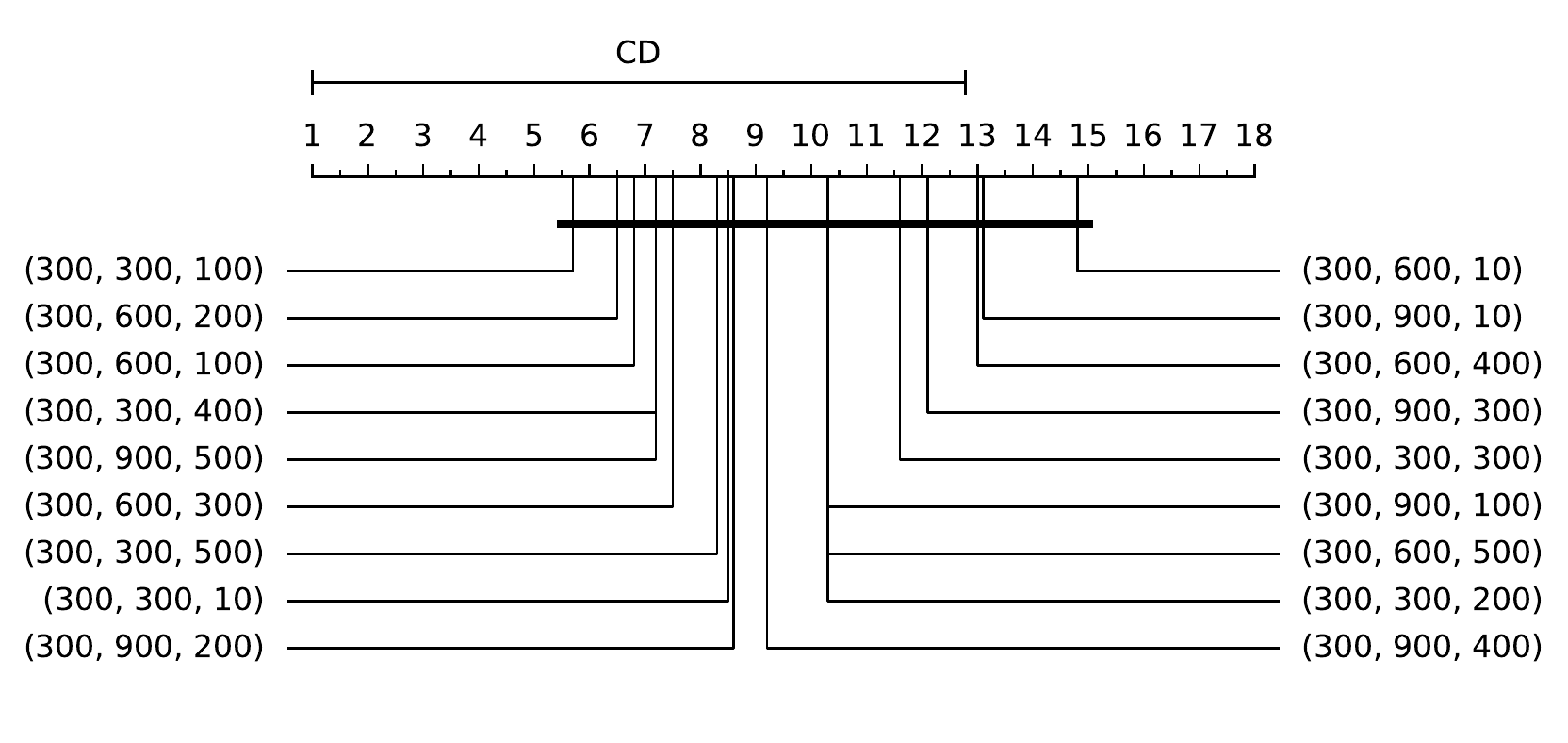}
  \end{center}
	\caption{
    The graph shows a critical difference (CD)
    among the various configuration of \sys.
    Each pair
    represents
    the three configurations; for example, (300, 600, 200) means
    $T_{prep} = 300$, $T_{focus} = 600$, $\theta_{init} = 200$.
    The CD diagram demonstrates that
    there is no statistical difference
    among the tested configurations
    (all configurations are grouped in one line).
  }
  \label{f:cd-combined-aimd-parameter-appendix}
\end{figure}
\begin{figure}[H]
  \centering
    \includegraphics[width=\columnwidth]{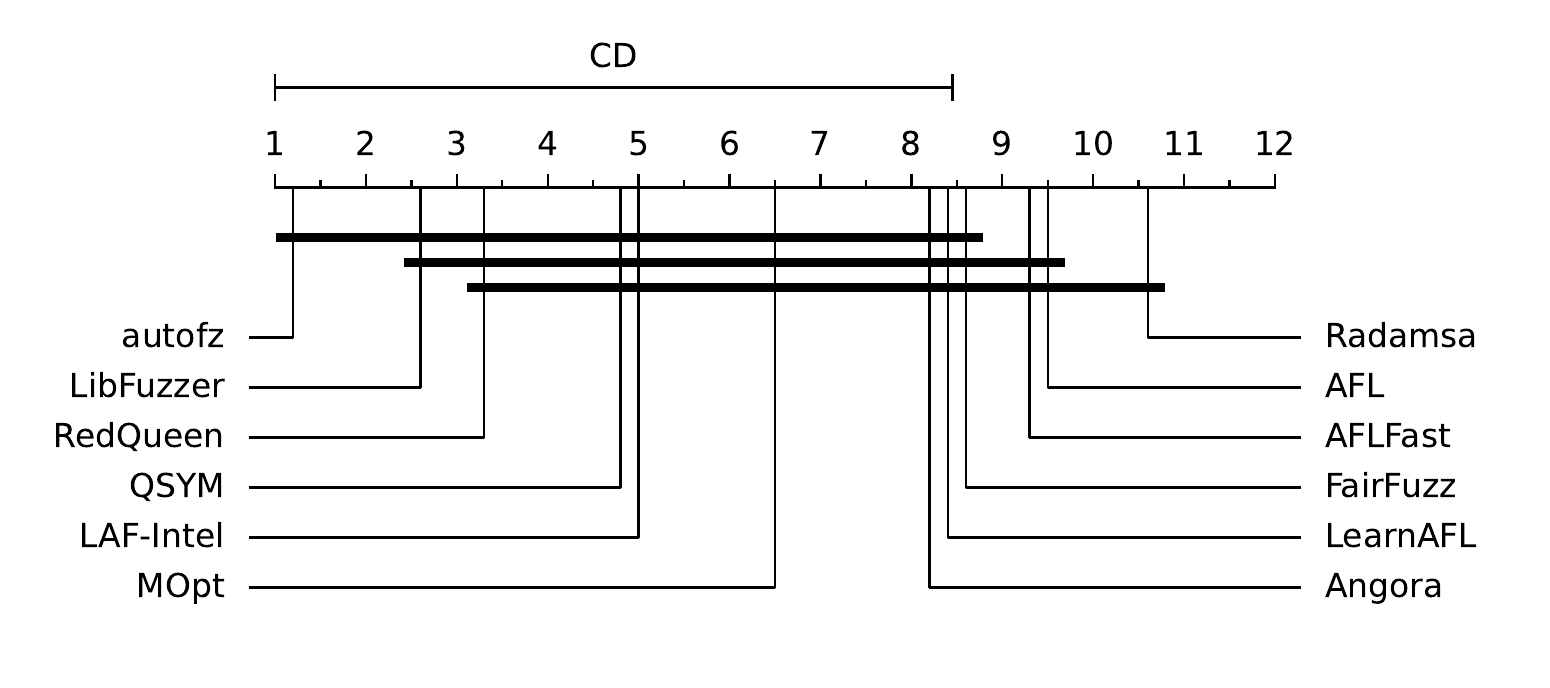}
  \caption{Critical Difference for the targets in \fts
   }
  \label{f:cd-fts}
\end{figure}
\end{minipage}
\hspace{0.05\textwidth}
\begin{minipage}{0.45\textwidth}
\begin{figure}[H]
\begin{subfigure}{\columnwidth}
    {\rowcolors{2}{}{gray!20}
  \begin{center}
  \resizebox{\columnwidth}{!}{
    \begin{tabular}{lrrrrr}
\hline
Round & Winner & $\pdiff$ & $\theta$ & $\prep$ & $\focus$  \\ \hline
   $1^*$ & \angora   & 857 & 100 & 30 & 570 \\ \hline
   $2^*$ & \redqueen & 234 & 200 & 150 & 450 \\ \hline
   3 & None      & 116 & 300 & 300 & 300 \\ \hline
   4 & None   & 48  & 150 & 300 & 300 \\ \hline
   $5^*$ & \redqueen & 92 & 75 & 300 & 300 \\ \hline
   6 & None  & 30 & 175 & 300 & 300 \\ \hline
   7 & None  & 22 & 87.5 & 300 & 300 \\ \hline
   8 & None  & 26 & 43.75 & 300 & 300 \\ \hline
   $9^*$ & \lafintel  & 71 & 21.88 & 90 & 510 \\ \hline
   10 & None  & 31 & 121.88 & 300 & 300 \\ \hline
   11 & None  & 21 & 60.94 & 300 & 300 \\ \hline
   12 & None  & 24 & 30.47 & 300 & 300 \\ \hline
   $13^*$ & \lafintel  & 27 & 15.23 & 300 & 300 \\ \hline
   14 & None  & 12 & 115.23 & 300 & 300  \\ \hline
   15 & None  & 40 & 57.62 & 300 & 300 \\ \hline
   \end{tabular}
  }
  \end{center}
  \footnotesize{* asterisk mark after the round numbers means that $Exit_{early}$ is true.}
  }
\end{subfigure}

\begin{subfigure}{\columnwidth}
  \includegraphics[width=\columnwidth]{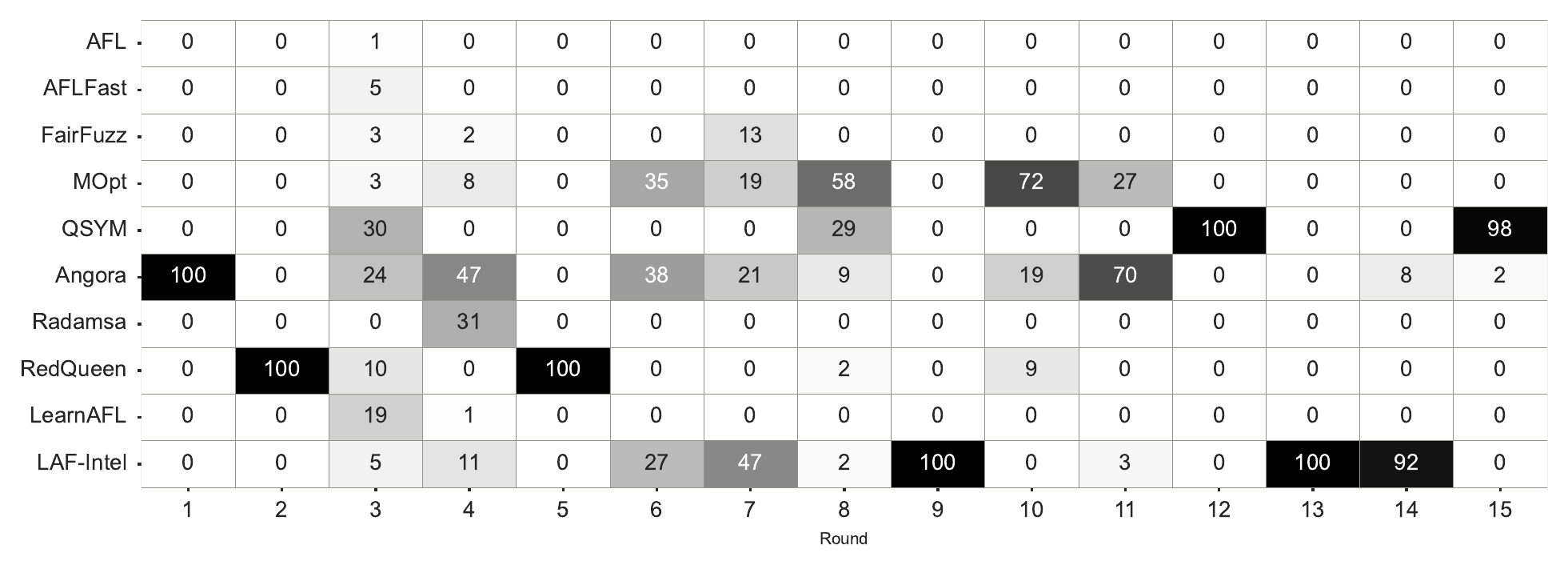}
\end{subfigure}
\caption{
Full result of \autoref{f:resource-distribute-exiv2}.
$\theta$ is rounded to two decimal places
}
\label{t:exiv2-round-winner-full}
\end{figure}
\end{minipage}

\subsection{Comparison of \sysn and collaborative fuzzing on \fts targets}
\label{s:fts-collab}
\makebox[\textwidth]{%
      \includegraphics[width=\textwidth]{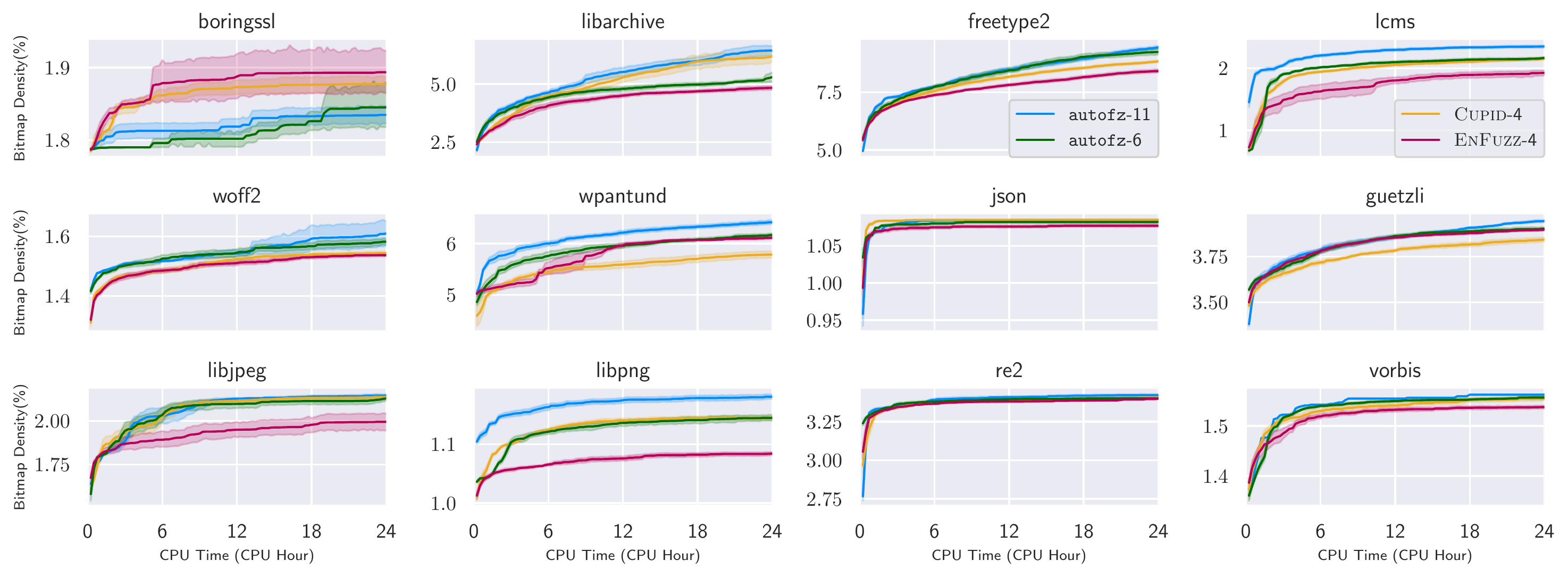}}
\captionof{figure}{
          Comparison among \sys, \enfuzz, and \cupid.
          Each line plot is depicted with an arithmetic mean
          and 80\% confidence interval for 10 times fuzzing runs.
          The coverage ratio is a percentage of branches explored by each fuzzer.
          X-axis is elapsed \textbf{\textit{CPU time}}.
          Below is the list of fuzzers used by different configurations:
         \textbf{\sys-6} = [\afl, \fairfuzz, \qsym, \aflfast, \libfuzzer, \radamsa],
          \textbf{\cupid-4} = [\afl, \fairfuzz, \qsym, \libfuzzer],
          \textbf{\enfuzz-4} = [\afl, \fairfuzz, \radamsa, \libfuzzer] which is reported to be the best in \cite{chen19:enfuz},
          \textbf{\sys-11} = [All baseline fuzzers described in~\autoref{s:eval}]
}
\label{f:autofuzz-enfuzz-cupid}

\end{minipage}

\clearpage

\begin{minipage}[H]{\textwidth}%

\begin{minipage}{0.45\textwidth}
    {
\rowcolors{2}{}{gray!20}
\begin{table}[H]
  \begin{center}
  \resizebox{\columnwidth}{!}{
    \begin{tabular}{lrrrr}
      \hline
 benchmark & \texttt{autofz}-11 & \texttt{autofz}-6 & \textsc{Cupid}-4 & \textsc{EnFuzz}-4 \\ \hline
boringssl & 1.85 & 1.85 & 1.88 & \textbf{1.89} \\ \hline
freetype2 & \textbf{9.88} & 9.26 & 8.85 & 8.43 \\ \hline
guetzli & \textbf{3.96} & 3.90 & 3.84 & 3.90 \\ \hline
json & \textbf{1.09} & 1.08 & \textbf{1.09} & 1.08 \\ \hline
lcms & \textbf{2.36} & 2.17 & 2.15 & 1.92 \\ \hline
libarchive & \textbf{6.74} & 5.30 & 6.23 & 4.83 \\ \hline
libjpeg & \textbf{2.15} & 2.13 & 2.14 & 1.99 \\ \hline
libpng & \textbf{1.18} & 1.14 & 1.14 & 1.08 \\ \hline
re2 & \textbf{3.42} & 3.40 & 3.41 & 3.40 \\ \hline
vorbis & \textbf{1.56} & \textbf{1.56} & 1.55 & 1.54 \\ \hline
woff2 & \textbf{1.62} & 1.58 & 1.54 & 1.54 \\ \hline
wpantund & \textbf{6.53} & 6.17 & 5.79 & 6.12 \\ \hline
    \hline 
    \textbf{SUM} & 42.35 & 39.55 & 39.61 & 37.72 \\ \hline
    \textbf{IMPROVE} & \textbf{+12.27\%}   & \textbf{+4.85\%}  & +5\% & - \\ \hline
     \end{tabular}
     }
  \end{center}
  \caption{Bitmap density of \sys, \cupid, and \enfuzz on \fts corresponding to
    \autoref{f:autofuzz-enfuzz-cupid}. The bitmap density is presented 
    as a percentage.}
  \label{t:cov-autofuzz-cupid-enfuzz}
\end{table}
}
\end{minipage}
\hspace{0.05\textwidth}
\begin{minipage}{0.45\textwidth}
    {\rowcolors{2}{}{gray!20}
\begin{table}[H]
    \begin{center}
\resizebox{\columnwidth}{!}{
  \begin{tabular}{lrr|rr}
    \hline
              & \multicolumn{2}{c}{\sys-11} & \multicolumn{2}{c}{\sys-6} \\ \hline
    benchmark & \cupid-4 & \enfuzz-4 & \cupid-4 & \enfuzz-4 \\ \hline
    boringssl & \textbf{< 0.01} & \textbf{0.03} & \textbf{< 0.01} & \textbf{0.02} \\ \hline
freetype2 & \textbf{< 0.01} & \textbf{< 0.01}  & \textbf{< 0.01} & \textbf{< 0.01} \\ \hline
guetzli & \textbf{< 0.01} & \textbf{< 0.01} & \textbf{< 0.01} & 0.76 \\ \hline
json & \textbf{< 0.01} & \textbf{< 0.01} & \textbf{0.02} & \textbf{0.02} \\ \hline
lcms & \textbf{< 0.01} & \textbf{< 0.01} & 0.76 & \textbf{< 0.01} \\ \hline
libarchive & \textbf{0.05} & \textbf{< 0.01} & 0.06 & \textbf{< 0.01} \\ \hline
libjpeg & 0.97 & \textbf{< 0.01} & 0.45 & \textbf{< 0.01} \\ \hline
libpng & \textbf{< 0.01} & \textbf{< 0.01} & 0.97 & \textbf{< 0.01} \\ \hline
re2 & 0.20 & \textbf{0.02}  & 0.32 & 1.00 \\ \hline
vorbis & 0.27 & 0.15 & 0.40 & \textbf{< 0.01} \\ \hline
woff2 & \textbf{0.03} & \textbf{0.02} & \textbf{< 0.01} & \textbf{< 0.01} \\ \hline
wpantund & \textbf{< 0.01} & \textbf{0.01} & \textbf{< 0.01} & 0.35 \\ \hline
  \end{tabular}
  }
\end{center}
  \caption{Mann-Whitney U Test evaluated on \fts
  based on the bitmap result
  presented in~\autoref{t:cov-autofuzz-cupid-enfuzz}.
  The left two columns compare \sys-11 with
  \cupid-4 and \enfuzz-4. 
  The right two columns compares \sys-6 and the others.
  The results showing a statistical difference are in \textbf{bold}.}
  \label{t:p-value-vs-enfuzz-cupid}
\end{table}
}
\end{minipage}

\subsection{Comparison of \sysn and collaborative fuzzing on \unifuzz targets}
    \begin{center}
    \begin{minipage}{0.45\textwidth}
        \rowcolors{3}{gray!20}{}
\begin{table}[H]
  \begin{center}
  \begin{tabular}{lrrr}
    \hline
    benchmark & \texttt{autofz}-10 & \texttt{autofz}-6 & \textsc{Cupid}-4 \& \\
    &&&\textsc{EnFuzz}-Q \\ \hline
    exiv2 & \textbf{19.67} & 14.56 & 10.18  \\ \hline
    ffmpeg & \textbf{99.52} & 99.27 & 44.24  \\ \hline
    imginfo & \textbf{8.32} & 6.45 & 3.77  \\ \hline
    infotocap & \textbf{6.64} & 6.32 & 5.27 \\ \hline
    mujs & \textbf{9.56} & 7.91 & 7.71  \\ \hline
    nm & \textbf{13.28} & 12.27 & 7.81  \\ \hline
    pdftotext & \textbf{24.20} & 23.20 & 19.67  \\ \hline
    tcpdump & \textbf{37.74} & 30.13 & 10.32  \\ \hline
    \hline 
    \textbf{SUM} & 218.91 & 200.11 & 108.97 \\ \hline
    \textbf{IMPROVE} & \textbf{+100.9\%}   & \textbf{+83.63\%}  &  -  \\ \hline
    \end{tabular}
  \end{center}
  \caption{Bitmap density of \sys, \cupid, and \enfuzz-Q (same as \cupid-4, omitted) on \unifuzz corresponding to
    \autoref{f:autofuzz-enfuzz-cupid-unifuzz}. The bitmap density is presented 
    as a percentage.}
  \label{t:cov-autofuzz-cupid-enfuzz-unifuzz}
\end{table}

    \end{minipage}
    \hspace{0.05\textwidth}
    \begin{minipage}{0.45\textwidth}
        \rowcolors{2}{}{gray!20}
\begin{table}[H]
  \begin{center}
  \begin{tabular}{lr|r}
    \hline
    & \multicolumn{1}{c}{\sys-10} & \multicolumn{1}{c}{\sys-6} \\ \hline
    benchmark & \textsc{Cupid}-4  & \textsc{Cupid}-4 \\ \hline
    exiv2 & \textbf{< 0.01} & \textbf{< 0.01}  \\ \hline
    ffmpeg & \textbf{< 0.01}   & \textbf{< 0.01} \\ \hline
    imginfo & \textbf{< 0.01}  & \textbf{< 0.01} \\ \hline
    infotocap & \textbf{< 0.01} & \textbf{< 0.01} \\ \hline
    mujs & \textbf{< 0.01}  & \textbf{0.05}  \\ \hline
    nm & \textbf{< 0.01}  & \textbf{< 0.01}  \\ \hline
    pdftotext & \textbf{0.04} & \textbf{< 0.01}  \\ \hline
    tcpdump & \textbf{< 0.01}   & \textbf{< 0.01} \\ \hline
  
  \end{tabular}
  \end{center}
  
    \caption{
    Mann-Whitney U Test evaluated on \unifuzz
    based on the bitmap result presented in \autoref{t:cov-autofuzz-cupid-enfuzz-unifuzz}.
    The left two columns compare \sys-10 with
    \cupid-4 and \enfuzz-Q (same as \cupid-4, omitted). 
    The right two columns compare \sys-6 and the others.
    The results showing statistical difference are in \textbf{bold}.
    }
  \label{t:p-value-vs-enfuzz-cupid-unifuzz}
\end{table}

    \end{minipage}
    \end{center}
\end{minipage}

\clearpage
\begin{minipage}{\textwidth}
    \subsection{Detail Coverage and Mann-Whitney U Test among \sysn and individual fuzzers}
    \begin{center}
    \begin{minipage}{\textwidth}
        {\rowcolors{2}{}{gray!20}
\begin{table}[H]
  \begin{center}
  \resizebox{\columnwidth}{!}{
    \begin{tabular}{lcccccccccccc}
\hline
benchmark & \sys & AFL & AFLFast & Angora & FairFuzz & LAF-Intel & LearnAFL & MOpt & QSYM & Radamsa & RQ & Lib \\ \hline
      exiv2 & \textbf{18.84} & 9.35 & 4.90 & 12.84 & 11.41 & 13.69 & 9.80 & 11.53 & 8.49 & 1.89 & 14.90 & - \\ \hline
      ffmpeg & \textbf{100.00} & 44.08 & 46.11 & 25.60 & 49.74 & 79.83 & \textbf{100.00} & 60.37 & 20.32 & 54.80 & 88.85  & - \\ \hline
      imginfo & \textbf{7.82} & 2.65 & 2.47 & 5.55 & 3.05 & 6.62 & 6.01 & 5.27 & 3.47 & 3.89 & 7.81  & - \\ \hline
      infotocap & \textbf{6.19} & 4.75 & 5.03 & 3.92 & 5.40 & 5.70 & 5.82 & 5.54 & 4.81 & 3.95 & 6.08  & - \\ \hline
      mujs & \textbf{8.88} & 7.20 & 7.25 & 4.55 & 8.17 & 7.65 & 7.89 & 7.77 & 7.24 & 7.11 & 8.22  & - \\ \hline
      nm & \textbf{12.19} & 6.14 & 5.80 & 6.58 & 6.92 & 10.59 & 9.14 & 11.01 & 7.85 & 4.45 & 10.73  & - \\ \hline
      pdftotext & 24.02 & 19.37 & 19.46 & 16.99 & 19.11 & 23.09 & 23.95 & 19.98 & 18.40 & 20.47 & \textbf{26.77}  & - \\ \hline
      tcpdump & \textbf{36.30} & 9.34 & 9.31 & 29.25 & 9.85 & 30.09 & 29.57 & 26.38 & 8.63 & 9.98 & 31.92  & - \\ \hline
      tiffsplit & \textbf{7.56} & 5.97 & 6.03 & 4.98 & 6.38 & 6.57 & 6.75 & 6.80 & 6.33 & 3.89 & 6.45  & - \\ \hline
      boringssl & \textbf{1.89} & 1.80 & 1.79 & 1.79 & 1.81 & 1.83 & 1.80 & 1.83 & 1.79 & 1.78 & 1.84 & 1.87 \\ \hline
      freetype2 & \textbf{9.87} & 6.31 & 6.55 & 5.61 & 7.09 & 7.36 & 7.37 & 7.47 & 7.81 & 5.84 & 9.73 & 8.38 \\ \hline
      libarchive & \textbf{6.91} & 2.36 & 2.13 & 3.31 & 2.87 & 4.78 & 3.11 & 3.07 & 4.31 & 1.67 & 5.72 & 5.21 \\ \hline
\end{tabular}
}
  \end{center}
  \caption{Bitmap density of \sys and individual fuzzers corresponding to
    \autoref{f:autofuzz-eval-best}. Each number represents bitmap density in percentage. 
    The best result of each benchmark suite is in \textbf{bold} in the table. 
    RQ and LIB indicate \redqueen and \libfuzzer, respectively.}
  \label{t:cov-vs-focus}
\end{table}
}
    \end{minipage}

    \begin{minipage}{\textwidth}
        {\rowcolors{2}{}{gray!20}
\begin{table}[H]
\resizebox{\columnwidth}{!}{
  \centering
  \begin{tabular}{lccccccccccc}
\hline
benchmark & AFL & AFLFast & Angora & FairFuzz & LAF-Intel & LearnAFL & MOpt & QSYM & Radamsa & RedQueen & LibFuzzer\\ \hline
    exiv2 & \textbf{< 0.01} & \textbf{< 0.01} & \textbf{< 0.01} & \textbf{< 0.01} & \textbf{< 0.01} & \textbf{< 0.01} & \textbf{< 0.01} & \textbf{< 0.01} & \textbf{< 0.01} & \textbf{< 0.01} & - \\ \hline
    ffmpeg & \textbf{< 0.01} & \textbf{< 0.01} & \textbf{< 0.01} & \textbf{0.01} & \textbf{< 0.01} & 0.29 & \textbf{< 0.01} & \textbf{< 0.01} & \textbf{< 0.01} & \textbf{< 0.01} & - \\ \hline
    imginfo & \textbf{< 0.01} & \textbf{< 0.01} & \textbf{< 0.01} & \textbf{< 0.01} & \textbf{< 0.01} & \textbf{< 0.01} & \textbf{< 0.01} & \textbf{< 0.01} & \textbf{< 0.01} & 0.24 & - \\ \hline
    infotocap & \textbf{< 0.01} & \textbf{< 0.01} & \textbf{< 0.01} & \textbf{< 0.01} & \textbf{0.03} & \textbf{0.03} & \textbf{< 0.01} & \textbf{< 0.01} & \textbf{< 0.01} & 0.09 & - \\ \hline
    mujs & \textbf{< 0.01} & \textbf{< 0.01} & \textbf{< 0.01} & \textbf{< 0.01} & \textbf{< 0.01} & \textbf{< 0.01} & \textbf{< 0.01} & \textbf{< 0.01} & \textbf{< 0.01} & \textbf{< 0.01} & - \\ \hline
    nm & \textbf{< 0.01} & \textbf{< 0.01} & \textbf{< 0.01} & \textbf{< 0.01} & \textbf{< 0.01} & \textbf{< 0.01} & \textbf{< 0.01} & \textbf{< 0.01} & \textbf{< 0.01} & \textbf{0.01} & - \\ \hline
    pdftotext & \textbf{< 0.01} & \textbf{< 0.01} & \textbf{< 0.01} & \textbf{< 0.01} & \textbf{< 0.01} & 0.60 & \textbf{< 0.01} & \textbf{< 0.01} & \textbf{< 0.01} & \textbf{< 0.01} & - \\ \hline
    tcpdump & \textbf{< 0.01} & \textbf{< 0.01} & \textbf{< 0.01} & \textbf{< 0.01} & \textbf{< 0.01} & \textbf{< 0.01} & \textbf{< 0.01} & \textbf{< 0.01} & \textbf{< 0.01} & \textbf{< 0.01} & - \\ \hline
    tiffsplit & \textbf{< 0.01} & \textbf{< 0.01} & \textbf{< 0.01} & \textbf{< 0.01} & \textbf{< 0.01} & \textbf{< 0.01} & \textbf{< 0.01} & \textbf{< 0.01} & \textbf{< 0.01} & \textbf{< 0.01} & - \\ \hline
    boringssl & \textbf{< 0.01} & \textbf{< 0.01} & \textbf{< 0.01} & \textbf{< 0.01} & \textbf{< 0.01} & \textbf{< 0.01} & \textbf{< 0.01} & \textbf{< 0.01} & \textbf{< 0.01} & \textbf{< 0.01} & 0.60 \\ \hline
freetype2 & \textbf{< 0.01} & \textbf{< 0.01} & \textbf{< 0.01} & \textbf{< 0.01} & \textbf{< 0.01} & \textbf{< 0.01} & \textbf{< 0.01} & \textbf{< 0.01} & \textbf{< 0.01} & 0.34 & \textbf{< 0.01} \\ \hline
libarchive & \textbf{< 0.01} & \textbf{< 0.01} & \textbf{< 0.01} & \textbf{< 0.01} & \textbf{< 0.01} & \textbf{< 0.01} & \textbf{< 0.01} & \textbf{< 0.01} & \textbf{< 0.01} & \textbf{< 0.01} & \textbf{< 0.01} \\ \hline
\end{tabular}
}
\caption{Mann-Whitney U Test results
evaluated based on the result presented in~\autoref{t:cov-vs-focus}.
A value less than 0.05 indicates
the result is statistically different
between \sys and the selected fuzzer.
The results showing statistical difference
are in \textbf{bold}.
}
  \label{t:p-value-vs-focus}
\end{table}
}

    \end{minipage}
    \end{center}

\subsection{Effects of the number of baseline fuzzers for \exiv2 and \nm}
\begin{figure}[H]
  \includegraphics[width=0.5\columnwidth]{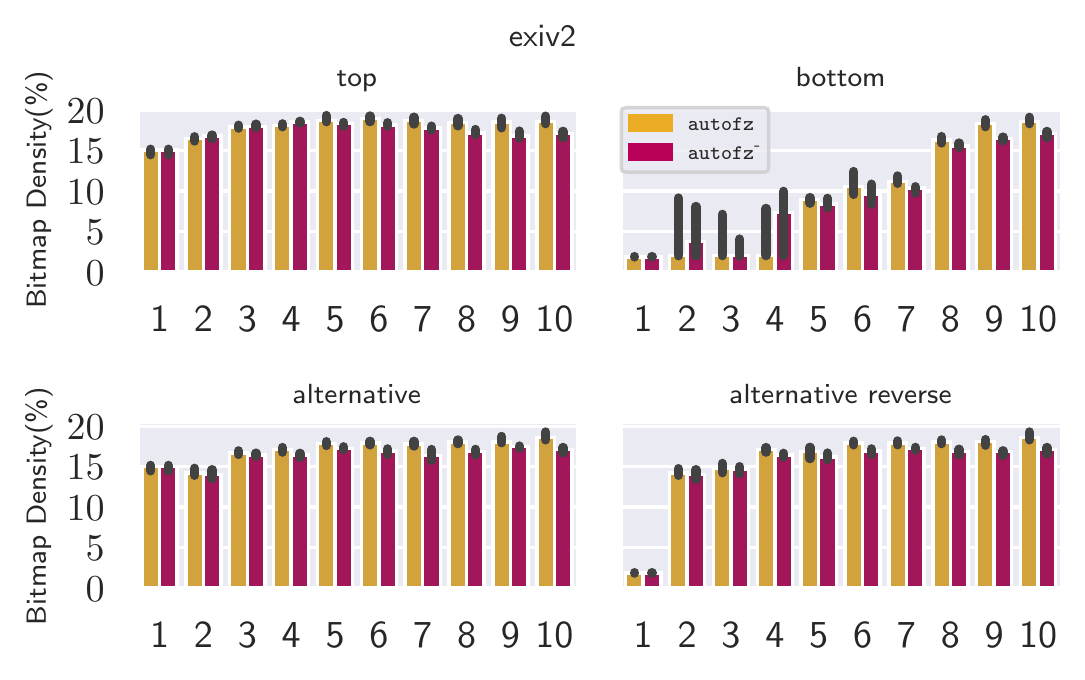}
  {\unskip\ \vrule\ }
  \includegraphics[width=0.5\columnwidth]{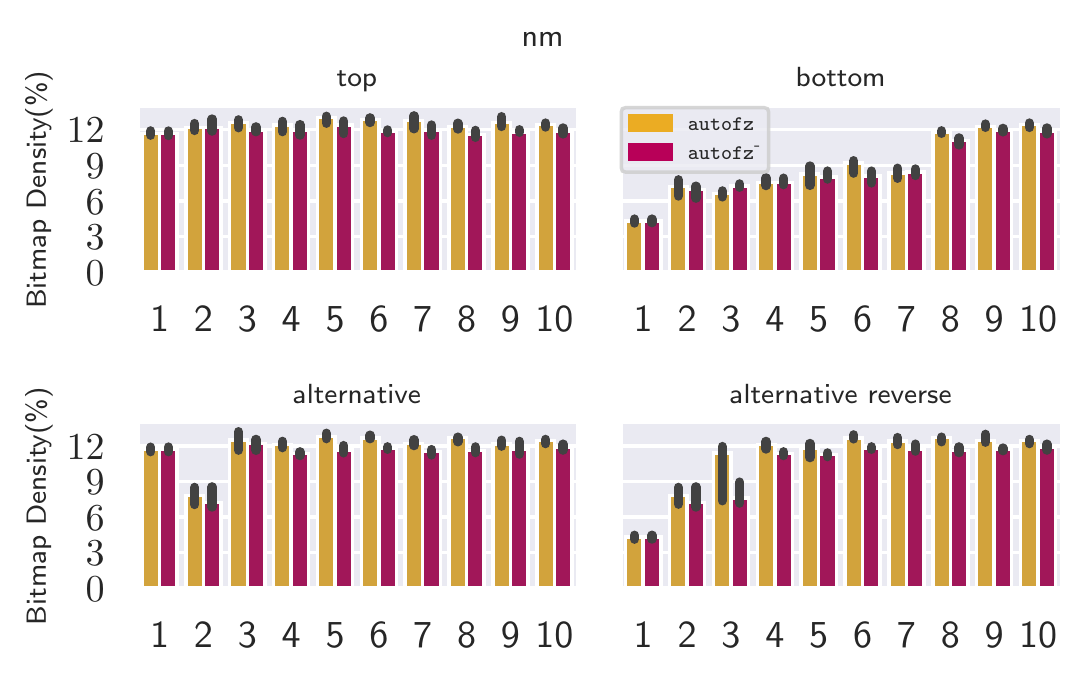}
  \caption{
  Evaluation of \sys and \sysm on
  different numbers of baseline fuzzers (1-10) for \cc{\exiv2} and \cc{nm}.
  }
  \label{f:fuzzer-num-exiv2}
\end{figure}
\end{minipage}

\clearpage
\begin{minipage}{\textwidth}
\subsection{Unique Bug Count among \sysn, individual fuzzers, and collaborative fuzzing}
\label{ss:bug-count}

\PP{Average number of unique bugs}
The unique bugs represent
all deduplicated bugs found by a specific fuzzer.
We compare the average number of unique bugs of \sys and others in 10 repetitions (\autoref{t:bug-vs-focus}, \autoref{t:bug-autofuzz-cupid-enfuzz-unifuzz}, and \autoref{t:bug-autofuzz-cupid-enfuzz}).
The result shows that \sys can outperform
individual fuzzers and collaborative fuzzing such as \enfuzz and \cupid
for bug finding.

\PP{Aggregated unique bug counts.}
In addition to the average number of unique bugs,
we also aggregate all unique bugs found by each fuzzer
in 10 repetitions.
We present the numbers
in three different categories per individual fuzzer:
the number of unique bugs,
the number of exclusive bugs, and
the number of unobserved bugs (\autoref{t:bug-vs-focus-detail},
\autoref{t:bug-autofuzz-cupid-enfuzz-unifuzz-detail}, and
\autoref{t:bug-autofuzz-cupid-enfuzz-unifuzz-detail}).
The exclusive bugs means the bugs that
cannot be found by other fuzzers, but only by this fuzzer.
The unobserved bugs are the bugs that
cannot be found by this fuzzer, but are reported by the others.

\PP {Bug report}
We could not find a new bug
because \unifuzz and \fts consist of old versions of programs, and
most of the bugs have been already reported and fixed.
We reproduced the found bug on the latest version of the targets
and confirmed that all of the bugs we found
are not present in the latest version.

{\rowcolors{2}{}{gray!20}
\begin{table}[H]
  \begin{center}
  \resizebox{\columnwidth}{!}{
    \begin{tabular}{lcccccccccccc}
\hline
benchmark & \sys & AFL & AFLFast & Angora & FairFuzz & LAF-Intel & LearnAFL & MOpt & QSYM & Radamsa & RQ & Lib \\ \hline
exiv2 & \textbf{13.10} & 2.50 & 1.50 & 7.10 & 3.10 & 5.60 & 2.70 & 4.30 & 0.00 & 0.10 & 6.40 & - \\ \hline
ffmpeg & 0.20 & 0.00 & 0.00 & 0.00 & 0.00 & \textbf{0.80} & 0.20 & 0.00 & 0.00 & 0.00 & \textbf{0.80} & - \\ \hline
imginfo & 0.80 & 0.00 & 0.00 & 0.00 & 0.20 & \textbf{1.00} & 0.60 & 0.30 & 0.00 & 0.00 & \textbf{1.00} & - \\ \hline
infotocap & \textbf{5.50} & 2.50 & 2.70 & 0.00 & 3.60 & 2.30 & 3.90 & 3.60 & 0.00 & 0.30 & 3.20 & - \\ \hline
mujs & 3.00 & 0.00 & 0.50 & 0.00 & 0.30 & 2.00 & 0.60 & 0.80 & 0.00 & \textbf{4.20} & 0.70 & - \\ \hline
nm & 0.50 & 0.00 & 0.00 & \textbf{0.70} & 0.00 & 0.10 & 0.00 & 0.10 & 0.00 & 0.00 & 0.20 & - \\ \hline
pdftotext & 3.70 & 1.50 & 1.60 & 1.20 & 1.40 & 6.80 & 6.50 & 1.90 & 0.00 & 2.30 & \textbf{9.10} & - \\ \hline
tcpdump & 1.90 & 0.00 & 0.00 & \textbf{2.30} & 0.00 & 1.50 & 1.00 & 1.50 & 0.00 & 0.00 & 0.70 & - \\ \hline
tiffsplit & \textbf{6.20} & 2.90 & 2.90 & 3.50 & 3.50 & 3.80 & 4.70 & 3.90 & 0.00 & 1.40 & 3.50 & - \\ \hline
boringssl & \textbf{0.40} & 0.00 & 0.00 & 0.00 & 0.00 & 0.00 & 0.00 & 0.00 & 0.00 & 0.00 & 0.00 & 0.20 \\ \hline
freetype2 & 0.00 & 0.00 & 0.00 & 0.00 & 0.00 & 0.00 & 0.00 & 0.00 & 0.00 & 0.00 & 0.00 & 0.00 \\ \hline
libarchive & 0.40 & 0.00 & 0.00 & 0.00 & 0.00 & 0.30 & 0.00 & 0.00 & 0.00 & 0.00 & \textbf{0.70} & 0.40 \\ \hline
\hline 
\textbf{SUM} & \textbf{35.70} & 9.40 & 9.20 & 14.80 & 12.10 & 24.20 & 20.20 & 16.40 & 0.00 & 8.30 & 26.30 & 0.60 \\ \hline
\end{tabular}
}
  \end{center}
  \caption{The average number of unique bugs found by \sys and individual fuzzers in 10 repetitions. 
  The RQ and LIB indicate \redqueen and \libfuzzer, respectively.}
  \label{t:bug-vs-focus}
\end{table}
}

{\rowcolors{2}{}{gray!20}
\begin{table}[H]
  \begin{center}
  \resizebox{\columnwidth}{!}{
    \begin{tabular}{lcccccccccccc}
\hline
benchmark & \sys & AFL & AFLFast & Angora & FairFuzz & LAF-Intel & LearnAFL & MOpt & QSYM & Radamsa & RQ & Lib \\ \hline
exiv2 & 31/4/11 & 8/0/34 & 8/0/34 & 20/3/22 & 9/0/33 & 23/3/19 & 13/1/29 & 16/0/26 & 0/0/42 & 1/0/41 & 17/1/25 & - \\ \hline
ffmpeg & 2/0/2 & 0/0/4 & 0/0/4 & 0/0/4 & 0/0/4 & 2/0/2 & 2/1/2 & 0/0/4 & 0/0/4 & 0/0/4 & 3/1/1 & - \\ \hline
imginfo & 1/0/0 & 0/0/1 & 0/0/1 & 0/0/1 & 1/0/0 & 1/0/0 & 1/0/0 & 1/0/0 & 0/0/1 & 0/0/1 & 1/0/0 & - \\ \hline
infotocap & 9/0/1 & 3/0/7 & 3/0/7 & 0/0/10 & 7/1/3 & 5/0/5 & 7/0/3 & 5/0/5 & 0/0/10 & 1/0/9 & 6/0/4 & - \\ \hline
mujs & 6/0/4 & 0/0/10 & 1/0/9 & 0/0/10 & 1/0/9 & 3/0/7 & 1/0/9 & 2/0/8 & 0/0/10 & 9/4/1 & 1/0/9 & - \\ \hline
nm & 2/1/8 & 0/0/10 & 0/0/10 & 6/4/4 & 0/0/10 & 1/0/9 & 0/0/10 & 1/1/9 & 0/0/10 & 0/0/10 & 2/2/8 & - \\ \hline
pdftotext & 11/1/28 & 2/0/37 & 3/0/36 & 2/0/37 & 2/0/37 & 16/2/23 & 19/10/20 & 4/0/35 & 0/0/39 & 4/0/35 & 23/11/16 & - \\ \hline
tcpdump & 6/0/9 & 0/0/15 & 0/0/15 & 11/5/4 & 0/0/15 & 7/2/8 & 3/0/12 & 3/0/12 & 0/0/15 & 0/0/15 & 1/0/14 & - \\ \hline
tiffsplit & 8/0/3 & 5/0/6 & 6/1/5 & 6/0/5 & 6/0/5 & 6/1/5 & 8/1/3 & 5/0/6 & 0/0/11 & 3/0/8 & 8/0/3 & - \\ \hline
boringssl & 1/0/0 & 0/0/1 & 0/0/1 & 0/0/1 & 0/0/1 & 0/0/1 & 0/0/1 & 0/0/1 & 0/0/1 & 0/0/1 & 0/0/1 & 1/0/0 \\ \hline
freetype2 & 0/0/0 & 0/0/0 & 0/0/0 & 0/0/0 & 0/0/0 & 0/0/0 & 0/0/0 & 0/0/0 & 0/0/0 & 0/0/0 & 0/0/0 & 0/0/0 \\ \hline
libarchive & 1/0/2 & 0/0/3 & 0/0/3 & 0/0/3 & 0/0/3 & 1/0/2 & 0/0/3 & 0/0/3 & 0/0/3 & 0/0/3 & 1/0/2 & 2/2/1 \\ \hline
\hline 
\textbf{SUM} & \textbf{78/6/68} & 18/0/128 & 21/1/125 & 45/12/101 & 26/1/120 & 65/8/81 & 54/13/92 & 37/1/109 & 0/0/146 & 18/4/128 & 63/15/83 & 3/2/1 \\ \hline
\end{tabular}
}
  \end{center}
  \caption{Unique bug count of \sys and individual fuzzers aggregated over 10 fuzzing repetitions, corresponding to \autoref{f:autofuzz-eval-best}. 
  Each number in \textbf{A/B/C} indicates the following.
    \textbf{A}: the number of unique bugs.
    \textbf{B}: the number of exclusive bugs.
    \textbf{C}: the number of unobserved bugs.
    }
  \label{t:bug-vs-focus-detail}
\end{table}
}
\end{minipage}

\clearpage

\begin{figure*}
\begin{minipage}{\textwidth}
\begin{minipage}{0.45\textwidth}
\rowcolors{3}{gray!20}{}
\begin{table}[H]
  \begin{center}
  \begin{tabular}{lrrr}
    \hline
    benchmark & \texttt{autofz}-10 & \texttt{autofz}-6 & \textsc{Cupid}-4 \& \\
    &&&\textsc{EnFuzz}-Q \\ \hline
    exiv2 & \textbf{16.70} & 5.80 & 2.70 \\ \hline
    ffmpeg & 0.00 & 0.00 & 0.00 \\ \hline
    imginfo & \textbf{0.70} & 0.10 & 0.00 \\ \hline
    infotocap & \textbf{5.90} & 4.70 & 2.40 \\ \hline
    mujs & 3.50 & \textbf{3.60} & 0.20 \\ \hline
    nm & \textbf{1.20} & 0.00 & 0.00 \\ \hline
    pdftotext & \textbf{4} & 3.80 & 1.10 \\ \hline
    tcpdump & \textbf{1.00} & 0.50 & 0.00 \\ \hline
    \hline 
    \textbf{SUM} & \textbf{33} & 18.5 &  6.4  \\ \hline
    \textbf{IMPROVE} & \textbf{+415\%}   & \textbf{+189\%}  &  -  \\ \hline
    \end{tabular}
  \end{center}
  \caption{Unique Bug Count of \sys, \cupid, and \enfuzz-Q (same as \cupid-4, omitted) on \unifuzz corresponding to
    \autoref{f:autofuzz-enfuzz-cupid-unifuzz}. Each number represents the average bug count across 10 fuzzing repetitions.}
  \label{t:bug-autofuzz-cupid-enfuzz-unifuzz}
\end{table}
\end{minipage}
\hspace{0.05\textwidth}
\begin{minipage}{0.45\textwidth}
\rowcolors{3}{gray!20}{}
\begin{table}[H]
  \begin{center}
  \begin{tabular}{lrrr}
    \hline
    benchmark & \texttt{autofz}-10 & \texttt{autofz}-6 & \textsc{Cupid}-4 \& \\
    &&&\textsc{EnFuzz}-Q \\ \hline
    exiv2 & 33/16/4 & 20/4/17 & 6/0/31 \\ \hline
    ffmpeg & 0/0/0 & 0/0/0 & 0/0/0 \\ \hline
    imginfo & 1/0/0 & 1/0/0 & 0/0/1 \\ \hline
    infotocap & 7/1/2 & 8/2/1 & 5/0/4 \\ \hline
    mujs & 6/1/2 & 7/2/1 & 1/0/7 \\ \hline
    nm & 8/8/0 & 0/0/8 & 0/0/8 \\ \hline
    pdftotext & 9/0/0 & 9/0/0 & 2/0/7 \\ \hline
    tcpdump & 5/4/0 & 1/0/4 & 0/0/5 \\ \hline
    \hline 
    \textbf{SUM} & \textbf{69/30/8} & 46/8/31 & 14/0/63  \\ \hline
    \textbf{IMPROVE} & \textbf{+392\%}   & \textbf{+228\%}  &  -  \\ \hline
    \end{tabular}
  \end{center}
  \caption{Aggregated unique bug count of \sys, \cupid, and \enfuzz-Q (same as \cupid-4, omitted) across 10 fuzzing repetitions on \unifuzz corresponding to
    \autoref{f:autofuzz-enfuzz-cupid-unifuzz}.  Please refer to \autoref{t:bug-vs-focus-detail} for the meaning of cells.}
  \label{t:bug-autofuzz-cupid-enfuzz-unifuzz-detail}
\end{table}
\end{minipage}

\begin{minipage}{0.45\textwidth}
{
\rowcolors{2}{}{gray!20}
\begin{table}[H]
  \begin{center}
  \resizebox{\columnwidth}{!}{
    \begin{tabular}{lrrrr}
      \hline
 benchmark & \texttt{autofz}-11 & \texttt{autofz}-6 & \textsc{Cupid}-4 & \textsc{EnFuzz}-4 \\ \hline
    boringssl & 0.00 & 0.00 & 0.20 & \textbf{0.40} \\ \hline
freetype2 & 0.00 & 0.00 & 0.00 & 0.00 \\ \hline
guetzli & \textbf{1.10} & \textbf{1.10} & 0.60 & \textbf{1.10} \\ \hline
json & \textbf{1.00} & \textbf{1.00} & \textbf{1.00} & \textbf{1.00} \\ \hline
lcms & 0.00 & 0.00 & 0.00 & 0.00 \\ \hline
libarchive & \textbf{0.40} & 0.00 & 0.00 & 0.00 \\ \hline
libjpeg & 0.00 & 0.00 & 0.00 & 0.00 \\ \hline
libpng & 0.00 & \textbf{0.10} & 0.00 & 0.00 \\ \hline
re2 & \textbf{0.30} & 0.00 & 0.10 & \textbf{0.30} \\ \hline
woff2 & \textbf{0.50} & 0.30 & 0.20 & 0.40 \\ \hline
vorbis & 0.00 & 0.00 & 0.00 & \textbf{0.10} \\ \hline
wpantund & 0.00 & 0.00 & 0.00 & 0.00 \\ \hline
    \hline 
    \textbf{SUM} & \textbf{3.3} & 2.5 & 2.1 & \textbf{3.3} \\ \hline
     \end{tabular}
     }
  \end{center}  
  \caption{Unique Bug Count of \sys, \cupid, and \enfuzz on \fts corresponding to
    \autoref{f:autofuzz-enfuzz-cupid}. Each number represents the average bug count across 10 fuzzing repetitions.}
  \label{t:bug-autofuzz-cupid-enfuzz}
\end{table}
}
\end{minipage}
\hspace{0.05\textwidth}
\begin{minipage}{0.45\textwidth}
{
\rowcolors{2}{}{gray!20}
\begin{table}[H]
  \begin{center}
  \resizebox{\columnwidth}{!}{
    \begin{tabular}{lrrrr}
      \hline
 benchmark & \texttt{autofz}-11 & \texttt{autofz}-6 & \textsc{Cupid}-4 & \textsc{EnFuzz}-4 \\ \hline
boringssl & 0/0/2 & 0/0/2 & 1/0/1 & 2/1/0 \\ \hline
freetype2 & 0/0/0 & 0/0/0 & 0/0/0 & 0/0/0 \\ \hline
guetzli & 2/0/0 & 2/0/0 & 1/0/1 & 2/0/0 \\ \hline
json & 1/0/0 & 1/0/0 & 1/0/0 & 1/0/0 \\ \hline
lcms & 0/0/0 & 0/0/0 & 0/0/0 & 0/0/0 \\ \hline
libarchive & 3/3/0 & 0/0/3 & 0/0/3 & 0/0/3 \\ \hline
libjpeg & 0/0/0 & 0/0/0 & 0/0/0 & 0/0/0 \\ \hline
libpng & 0/0/1 & 1/1/0 & 0/0/1 & 0/0/1 \\ \hline
re2 & 1/0/0 & 0/0/1 & 1/0/0 & 1/0/0 \\ \hline
woff2 & 1/0/0 & 1/0/0 & 1/0/0 & 1/0/0 \\ \hline
vorbis & 0/0/1 & 0/0/1 & 0/0/1 & 1/1/0 \\ \hline
wpantund & 0/0/0 & 0/0/0 & 0/0/0 & 0/0/0 \\ \hline
    \hline 
    \textbf{SUM} & \textbf{8/3/4} & 5/1/7 & 5/0/7 & \textbf{8/2/4} \\ \hline
     \end{tabular}
     }
  \end{center}
  \caption{Aggregated unique bug count of \sys, \cupid, and \enfuzz across 10 fuzzing repetitions on \fts corresponding to
    \autoref{f:autofuzz-enfuzz-cupid}. Please refer to \autoref{t:bug-vs-focus-detail} for the meaning of cells.}
  \label{t:bug-autofuzz-cupid-enfuzz-detail}
\end{table}
}
\end{minipage}
\end{minipage}
\end{figure*}
\clearpage
\begin{figure*}
\begin{minipage}{\textwidth}
\subsection{Different metric other than \afl bitmap}
To facilitate comparisons between \sys and
existing results in the literature,
we represent the results in three different metrics:
edge coverage (\autoref{f:autofuzz-eval-edge}),
line coverage (\autoref{f:autofuzz-eval-line}),
and queue size (\autoref{f:autofuzz-eval-queue}).
We adopt the same configuration of \sys
used in \autoref{f:autofuzz-eval-best}.
Each line plot is depicted with
an arithmetic mean and 80\% confidence interval
for 10 times fuzzing executions.
\begin{figure}[H]
  \begin{center}
    {
    \fontsize{6pt}{6pt}\selectfont
     \def\svgwidth{\textwidth}
     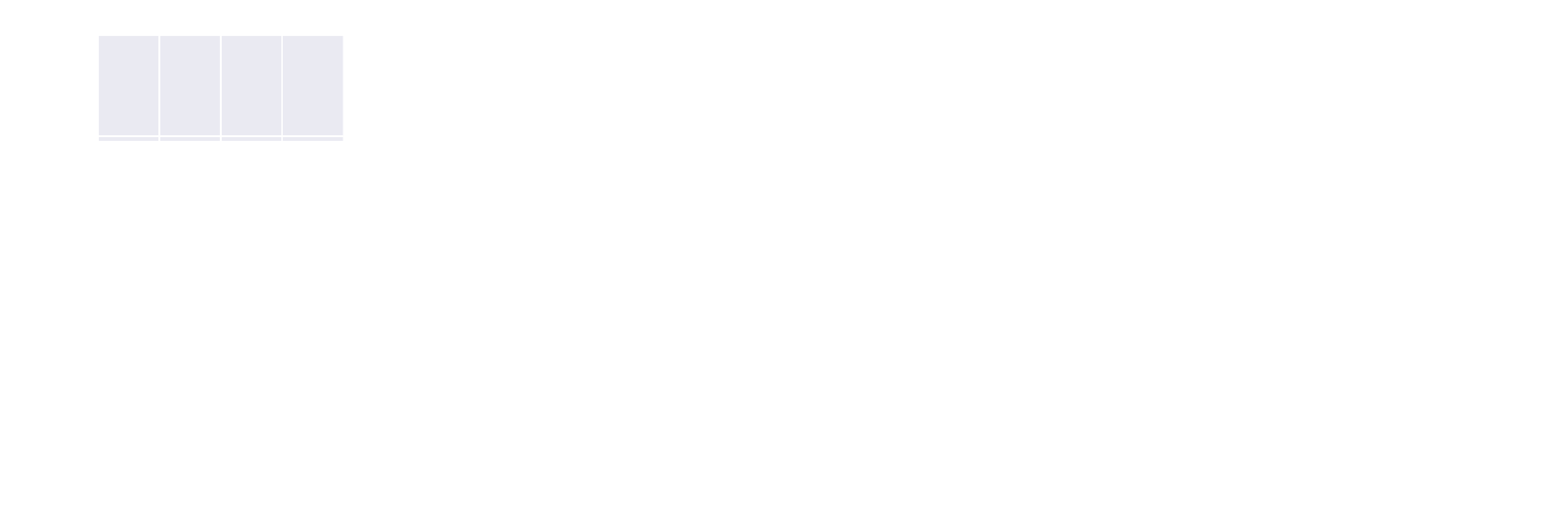
    }
  \end{center}
  \vspace{-0.45cm}
  \caption{Evaluation of \sys on \unifuzz and \fts in terms of \textbf{edge coverage.}}
\label{f:autofuzz-eval-edge}
\end{figure}
\vspace{-0.5cm}
\end{minipage}
\end{figure*}

\end{document}